\documentclass{article}
\usepackage{amsfonts}

\usepackage{graphicx,color}
\usepackage{amsmath}

\newenvironment{proof}[1][Proof]{\textbf{#1.} }{\ \rule{0.5em}{0.5em}}

\definecolor{green}{rgb}{0.00,0.50,0.00}

\input{tcilatex}

\begin{document}

\title{On the Stochastic Limit of Quantum Field Theory}
\author{L. Accardi, J.Gough and Y.G. Lu}
\date{}
\maketitle

\begin{abstract}
The weak coupling limit for a quantum system, with discrete energy spectrum,
coupled to a bose reservoir with the most general linear interaction is
considered: under this limit we have a quantum noise processes substituting
for the field. We obtain a limiting evolution unitary on the system and
noise space which, when reduced to the sytem's degrees of freedom, provide
the master and Langevin equations that are postulated on heuristic grounds
by physicists. In addition we give a concrete application of our results by
deriving the evolution of an atomic system interacting with the
electrodynamic field without recourse to either rotating wave or dipole
approximations. \ 
\end{abstract}

\section{\textbf{QUANTUM THEORY OF DAMPING}}

Irreversible quantum evolutions now play a fundamental role in many areas of
physics especially quantum optics. A large body of physical literature has
been built up around the problem of describing in a stochastic model the
effect of a source of quantum noise on a given quantum mechanical system,
emphasising the quantum stochastic properties of the source of the quantum
noise. However, quite generally, the classical approach to the stochasticity
which is taken ignores the essential quantum nature of the problem. In this
paper we wish to develop the weak coupling approach \cite{1} to quantum
damping. In a sense the noise feilds used to model physical sources are
still quantum in nature, as will be explained later, and because they are
arrive at from a well defined physical scaling limit do not require us to
put in the desired features of the noise by hand.

The derivations of the quantum master equations and quantum Langevin
equations for open quantum systems which are present in the current physics
literature are well motivated from the physical point of view, cf. \cite{2},
however mathematically imprecise. The heuristic procedure to render the
reservoir, to which the system is coupled, into a source of quantum noise
via some \textit{markovian approximation} is generally ill-defined and
reliant on second order perturbation theory.

On the other hand, the weak coupling limit for an open quantum system gives
a device for obtaining irreversible evolutions. Mathematically rigorous
derivations of the Langevin and master equations along these lines have been
given for certain specific models by Pul\'{e} \cite{3} and Davies \cite{4}.
Other attempts have been made to approximate a quantum reservoir by fitting
Quantum Brownian Motions (QBMs), cf. \cite{5},\cite{6} and \cite{7}.

In a long series of papers [\cite{1},\cite{8},\cite{9},\cite{10},\cite{11}],
Accardi, Frigerio and Lu have developed their approach whereby a quantum
reservoir can be reduced to a quantum stochastic noise source via a weak
coupling limit procedure. The theory is mathematically rigorous while at the
same time applicable to the wide range of phenomena considered by physicists
and gives a precise description of the reservoir as a quantum noise source.
The main mathematical device for establishing convergence of the reservoir
feilds is a quantum central limit theorem; that is a central limit theorem
for non-commuting observables: the theory of non-commutative probability, or 
\textit{Quantum Probability} as it is more correctly known, affords the
necessary mathematical framework to interpret the limit process. Indeed the
Fock space descriptions of a bose reservoir and a quantum stochastic
processes are so similar in nature, it is not surprising, with hindsight,
that the former may be reduced to the latter following some limiting
procedure.

Our objective in this paper is to review the results of the programme of
Accardi, Frigerio and Lu so far and to extend their theory so as to deal
with the most general interaction between a system and a noise source
encountered in physical theories. This we do and show that the energy
shifts, linewidths, master equations and Langevin equations, arising for the
system as a result of its coupling to the noise field, concur with those
obtain by earlier researchers \cite{2}. However we have, in addition, also a
quantum stochastic description of the reservoir noise fields themselves. As
a concrete application of our theory we consider the particular case where a
quantum electrodynamical field acts as reservoir, however we stress that
this is only one of the many applications of our theory.

We shall discuss only minimal coupling interacions, that is interactions
linear in the creation/ annihilation operators for the reservoir. In a later
paper we discuss how to treat the situations where the interaction is of
polynomial type.

\subsection{\textbf{Open Quantum Systems}}

We consider a system (S) coupled to a reservoir (R). The system (S) is to be
a quantum mechanical: its state space shall be a separable Hilbert space $%
\mathcal{H}_{S}$. The reservoir, on the other hand, is comprised of one or
indeed several quantum fields, and so has infinitely many degrees of
freedom. We shall consider a bosonic reservoir; the state space for $(R)$ is
the bosonic Fock space $\mathcal{H}_{R}$ over a separable Hilbert space $%
\mathcal{H}_{R}^{1}$; in standard notation we write $\mathcal{H}_{R}=\Gamma
_{B}(\mathcal{H}_{R}^{1})$. $\mathcal{H}_{R}^{1}$ is again to be a separable
Hilbert space and may quite generally describe not only one but several
individual species of particle in the reservoir. For instance, consider
several species of particle $P_{1},P_{2},P_{3},...$ in the reservoir and
suppose that $\mathcal{H}_{R}^{1}=\oplus _{j}\mathcal{H}_{P_{j}}^{1}$, where 
$\mathcal{H}_{P_{j}}^{1}$ is the state space for particle type $P_{j}$, then 
\begin{equation}
\mathcal{H}_{R}=\Gamma _{B}(\mathcal{H}_{R}^{1})=\Gamma _{B}(\oplus _{j}%
\mathcal{H}_{P_{j}}^{1})=\otimes _{j}\Gamma _{B}(\mathcal{H}_{P_{j}}^{1}). 
\TCItag{$1.1:1$}
\end{equation}
The space $\mathcal{H}_{R}^{1}$ is referred to as the (combined) one
particle state space for the reservoir. The overall state space for the
combined system and reservoir is $\mathcal{H}_{S}\otimes \mathcal{H}_{R}$.
The vacuum vector of the reservoir space will be denoted throughout as $\Psi
_{R}$.

In the following we consider only bosonic species in the reservoir, however
it is also possible to work with fermions \cite{9}. The dynamics of the
combined system and reservoir is governed by the formal Hamiltonian $%
H^{(\lambda )}$ which we may write as 
\begin{equation}
H^{(\lambda )}=H^{(0)}+\lambda H_{I};  \TCItag{$1.1:2$}
\end{equation}
that is, as the sum of a free Hamiltonian $H^{(0)}$ and an interaction $H_{I}
$, with $\lambda $ a real coupling parameter. $H^{(0)}$ is to be expressible
as 
\begin{equation}
H^{(0)}=H_{S}\otimes 1_{R}+1_{S}\otimes H_{R},  \TCItag{$1.1:3$}
\end{equation}
where $H_{S}$ and $H_{R}$ are self-adjoint operators on the spaces $\mathcal{%
H}_{S}$ and $\mathcal{H}_{R}$ respectively. For each $\lambda $, we consider
the unitary operator $V_{t}^{(\lambda )}$ on $\mathcal{H}_{S}\otimes 
\mathcal{H}_{R}$ defined by 
\begin{equation}
V_{t}^{(\lambda )}=\exp {\dfrac{t}{i\hbar }}H^{(\lambda )}.  \TCItag{$1.1:4$}
\end{equation}
This gives the time evolution under $H^{(\lambda )}$. A standard device in
pertubation theory is to transform to the interaction picture; this involves
introducing the operator 
\begin{equation}
U_{t}^{(\lambda )}=V_{t}^{(0)\dagger }V_{t}^{(\lambda )}.  \TCItag{$1.1:5$}
\end{equation}
$U_{t}^{(\lambda )}$ is a unitary operator on $\mathcal{H}_{S}\otimes 
\mathcal{H}_{R}$ called the wave operator at time $t$. We note that $%
\{U_{t}^{(\lambda )}:t\in \mathbb{R}\}$ is a left $v_{t}^{(0)}$-cocycle,
that is it satisfies the relation 
\begin{equation}
U_{t+s}^{(\lambda )}=v_{t}^{(0)}(U_{s}^{(\lambda )})U_{t}^{(\lambda )}. 
\TCItag{$1.1:6$}
\end{equation}
We define the time-evolutes for $X\in \mathcal{B}(\mathcal{H}_{S})\otimes 
\mathcal{B}(\mathcal{H}_{R})$ as 
\begin{equation*}
v_{t}^{(\lambda )}(X)=V_{t}^{(\lambda )\dagger }XV_{t}^{(\lambda )},
\end{equation*}
\begin{equation}
u_{t}^{(\lambda )}(X)=U_{t}^{(\lambda )\dagger }XU_{t}^{(\lambda )}. 
\TCItag{$1.1:7$}
\end{equation}
The Schr\"{o}dinger equation for the time evolutions are 
\begin{equation*}
{\dfrac{\partial }{\partial t}}V_{t}^{(\lambda )}={\dfrac{1}{i\hbar }}%
H^{(\lambda )}V_{t}^{(\lambda )}
\end{equation*}
and 
\begin{equation}
{\dfrac{\partial }{\partial t}}U_{t}^{(\lambda )}={\dfrac{\lambda }{i\hbar }}%
v_{t}^{(0)}(H_{I})U_{t}^{(\lambda )}.  \TCItag{$1.1:8$}
\end{equation}
With these we deduce the associated Heisenberg equations 
\begin{equation*}
{\dfrac{\partial }{\partial t}}v_{t}^{(\lambda )}(X)={\dfrac{1}{i\hbar }}%
[v_{t}^{(\lambda )}(X),v_{t}^{(\lambda )}(H^{(\lambda )})]={\dfrac{1}{i\hbar 
}}v_{t}^{(\lambda )}([X,H^{(\lambda )}])
\end{equation*}
and 
\begin{equation}
{\dfrac{\partial }{\partial t}}u_{t}^{(\lambda )}(X)={\dfrac{\lambda }{%
i\hbar }}u_{t}^{(\lambda )}([X,v_{t}^{(0)}(H_{I})]).  \TCItag{$1.1:9$}
\end{equation}
From (8) we obtain the integral equation 
\begin{equation}
U_{t}^{(\lambda )}=1+{\dfrac{\lambda }{i\hbar }}\int_{0}^{t}ds%
\,v_{s}^{(0)}(H_{I})\,U_{s}^{(\lambda )},  \TCItag{$1.1:10$}
\end{equation}
and consequently the iterated B\/orn series 
\begin{equation}
U_{t}^{(\lambda )}=1+\sum_{n=1}^{\infty }({\dfrac{\lambda }{i\hbar }}^n
\int_0^t dt_1 ...\int_{0}^{t_{n-1}}dt_{n}\,v_{t_{1}}^{(0)}(H_{I})...v_{t_{n}}^{(0)}(H_{I}).
\TCItag{$1.1:11$}
\end{equation}
This may be resummed as 
\begin{equation}
U_{t}^{(\lambda )}=\mathcal{T}\exp \{{\dfrac{\lambda }{i\hbar }}%
\int_{0}^{t}ds\,v_{s}^{(0)}(H_{I})\}.  \TCItag{$1.1:12$}
\end{equation}
where $\mathcal{T}$ denotes time ordering.

Before we continue, we must say more about how to interpret the formal sum
of $H^{(0)}$ and $H_{I}$. Firstly we assume that $H^{(0)}$ and $H_{I}$ are
self-adjoint operators on $\mathcal{H}_{S}\otimes \mathcal{H}_{R}$. We shall
assume that, for sufficiently small $\lambda $ and bounded $t$, the iterated
series (11) is uniformly convergent and is bounded on $\mathcal{H}_{S}%
\underline{\otimes }\mathcal{E}(\mathcal{H}_{R}^{1})$, the algebraic tensor
product of $\mathcal{H}_{S}$ nd the set of exponential vectors. From the
cocycle relation (1.1:6) we have that if we define the unitary operator $%
V_{t}^{(\lambda )}$ by 
\begin{equation}
V_{t}^{(\lambda )}=V_{t}^{(0)}U_{t}^{(\lambda )}  \TCItag{$1.1:13$}
\end{equation}
then $\{V_{t}^{(\lambda )}:t\in \mathbb{R}\}$ gives a strongly continuous
unitary group whose generator ${\dfrac{1}{\hbar }}H^{(\lambda )}$ is
formally given as $H^{(\lambda )}=H^{(0)}+\lambda H_{I}$. The time evolution
in the Heisenberg picture is then given by 
\begin{equation}
v_{t}^{(\lambda )}(X)=u_{t}^{(\lambda )}(v_{t}^{(0)}(X)).  \TCItag{$1.1:14$}
\end{equation}

\subsection{The Free Evolution}

For $\mathcal{H}_{R}^{1}=L^{2}(\mathbb{R}^{3})$ (momentum space), we have
creation/ annihilation operators $a^{\sharp }(\mathbf{k})$ satisfying 
\begin{equation}
\lbrack a(\mathbf{k}),a(\mathbf{k^{\prime }})^{\dagger }]=\delta (\mathbf{k}-%
\mathbf{k^{\prime }}),\,[a(\mathbf{k})^{\dagger },a^{\dagger }(\mathbf{%
k^{\prime }})]=0,\,[a(\mathbf{k}),a(\mathbf{k^{\prime }})]=0.  \TCItag{$1.2:1$}
\end{equation}
The creation/ annihilation fields on $\mathcal{H}_{R}$ are defined, for $%
g\in \mathcal{H}_{R}^{1}$, by 
\begin{equation}
A^{\dagger }(g)=\int d^{3}k\,g(\mathbf{k})a^{\dagger }(\mathbf{k}%
),\,A(g)=\int d^{3}k\,\overline{g}(\mathbf{k})a(\mathbf{k}).  \TCItag{$1.2:2$}
\end{equation}

From (16) we obtain the canonical commutation relations (CCR):

$[A(h),A^{\dagger }(f)]=$ $\langle h,f\rangle $, while $[A(h),A(f)]=0.$ We
take $H_{R}$ to be the second quantization of an operator $H_{R}^{1}$ on $%
\mathcal{H}_{R}^{1}$ given by 
\begin{equation}
(H_{R}^{1}f)(k)=\hbar \omega (k)f(k).  \TCItag{$1.2:3$}
\end{equation}
$H_{R}$ may then be expressed as 
\begin{equation}
H_{R}=\int dk\,\hbar \omega (k)\,a^{\dagger }(k)a(k).  \TCItag{$1.2:4$}
\end{equation}
Note that $v_{t}^{(0)}(1_{S}\otimes A^{\sharp }(g))=1_{S}\otimes A^{\sharp
}(S_{t}g);$ where we have introduced the unitary operator $S_{t}$ on $%
\mathcal{H}_{R}^{1}$ given by 
\begin{equation}
S_{t}=\exp \{-{\dfrac{t}{i\hbar }}H_{R}^{1}\};  \TCItag{$1.2:5$}
\end{equation}
that is $(S_{t}f)(k)=e^{i\omega (k)t}f(k)$.

\subsection{\textbf{The standard approach to the quantum Langevin equation}}

Consider an interaction of the type 
\begin{equation}
H_{I}=i\hbar \{D\otimes A^{\dagger }(g)-D^{\dagger }\otimes A(g)\}. 
\TCItag{$1.3:1$}
\end{equation}
where $D\in \mathcal{B}(\mathcal{H}_{S})$ has a harmonic free-evolution;
that is ${\dfrac{1}{i\hbar }}[D,H_{S}]=-i\omega D$. Then we have 
\begin{equation}
v_{t}^{(0)}(H_{I})=i\hbar \{D\otimes A^{\dagger }(S_{t}^{\omega
}g)-D^{\dagger }\otimes A(S_{t}^{\omega }g)\},  \TCItag{$1.3:2$}
\end{equation}
where 
\begin{equation}
S_{t}^{\omega }=e^{-i\omega t}S_{t},  \TCItag{$1.3:3$}
\end{equation}
that is,$(S_{t}^{\omega }f)(\mathbf{k})=e^{i(\omega (\mathbf{k})-\omega )t}f(%
\mathbf{k}).$ Now write $X_{t}=u_{t}^{(\lambda )}(X\otimes 1_{R})$, for $%
X\in \mathcal{B}(\mathcal{H}_{S})$, then from (9) we have 
\begin{equation}
{\dfrac{\partial X_{t}}{\partial t}}=\lambda \{A_{t}^{\dagger
}(S_{t}^{\omega }g)[X,D]_{t}-[X,D^{\dagger }]_{t}A_{t}(S_{t}^{\omega }g)\}, \TCItag{$1.3:4$}
\end{equation}
where $A_{t}^{\sharp }(f)=u_{t}^{(\lambda )}(1_{S}\otimes A^{\sharp }(f))$.

Now one can show that $A_{t}(f)=\lambda \int_{0}^{\infty }ds\,D_{s}\langle
f,S_{s}^{\omega }g\rangle +1_{S}\otimes A(f)$, so substituting back into
(23) gives 
\begin{equation}
{\dfrac{\partial X_{t}}{\partial t}} =\lambda ^{2}\int_{0}^{\infty
}\{D_{s}^{\dagger }\phi ^{\omega }(s-t)[X,D]_{t}-[X,D^{\dagger }]_{t}\phi
^{\omega }(t-s)D_{s}\}  +\lambda \{\xi _{t}^{\omega }[X,D]_{t}-[X,D^{\dagger }]_{t}\xi
_{t}^{\omega \dagger }\}, % \TCItag{$1.3:5$}
\end{equation}
where $\phi ^{\omega }(t)=\langle g,S_{-t}^{\omega }g\rangle =\int d^{3}k|g(%
\mathbf{k})|^{2}e^{-i(\omega (\mathbf{k})-\omega )t}$ and $\xi ^{\omega
}(t)=1_{S}\otimes A(S_{t}^{\omega }g).$ In standard terminology $\phi (t)$
is called the memory function and $\xi _{t}^{\omega }$ the fluctuating
quantum force [2,17] or input field [18], albeit in the interaction picture.
One notes that, in the vacuum state, $\xi _{t}^{\omega }$ is gaussian
distributed and all first and second moments vanish except the two-point
function 
\begin{equation}
\langle \Psi _{R},\xi _{t}^{\omega }\xi _{t}^{\omega \dagger }\Psi
_{R}\rangle =\langle S_{t}^{\omega }g,S_{s}^{\omega }g\rangle =\phi ^{\omega
}(t-s).  \TCItag{$1.3:6$}
\end{equation}
The standard approach taken at this juncture is to introduce the so-called 
\textit{first Markov approximation}. Here, for example, one takes $\mathcal{H%
}_{R}^{1}=L^{2}(\mathbb{R})$, $g=\sqrt{\dfrac{\kappa }{2\pi }}$ (constant)
and $\omega (k)=k$. Then 
\begin{equation}
\phi ^{\omega }(t)=\int_{-\infty }^{\infty }dk{\dfrac{\kappa }{2\pi }}%
e^{i(k+\omega )t}=\kappa e^{i\omega t}\delta (t).  \TCItag{$1.3:7$}
\end{equation}
There are, however, several important objections to be made to this
approach. Firstly any physical details specific to the reservoir must be put
in by hand. Secondly, the spectrum of $H_{R}^{1}$ is here unbounded below,
this is necessary to produce the delta function correlation of white noise.
From a physical point of view this is unacceptable as $H_{R}^{1}$ must be
bounded below for stability. Finally, the fact that the frequency spectrum $%
\omega (k)=k$ is unbounded below precludes any possibility of dropping the
rotating wave approximation.

\subsection{\textbf{The Weak Coupling Limit}}

We now describe the ideas behind the weak coupling limit in the simplest
situation where we have taken a dipole and rotating wave approximation. We
define the following reservoir operator 
\begin{equation}
B_{t}^{(\omega ,\lambda )}(g)=A(\lambda \int_{0}^{t}dt_{1}S_{t_{1}}^{\omega
}g)=\lambda \int_{M}\mu (dk)\int_{0}^{t}dt_{1}e^{-i(\omega (k)-\omega )t_{1}}%
\overline{g}(k)a(k).  \TCItag{$1.4:1$}
\end{equation}
Calculating the two-point vacuum expectations gives 
\begin{equation*}
\langle \Psi _{R}(0),B_{t}^{(\omega ,\lambda )}(g)B_{s}^{(\omega ,\lambda
)\dagger }(f)\Psi _{R}(0)\rangle =\lambda
^{2}\int_{0}^{t}dt_{1}\int_{0}^{s}ds_{1}\langle S_{t_{1}}^{\omega
}g,S_{s_{1}}^{\omega }f\rangle
\end{equation*}
\begin{equation}
\equiv \int_{0}^{\lambda ^{2}t}du\int_{u/\lambda ^{2}-s}^{u/\lambda
^{2}}d\tau \langle S_{\tau }^{\omega }g,f\rangle ,  \TCItag{$1.4:2$}
\end{equation}
where we have substituted $u=\lambda ^{2}t_{1}$ and $\tau =t_{1}-s_{1}$.

In order to obtain a non-trivial two-point function in the limit $\lambda
\rightarrow 0$ we must rescale time as 
\begin{equation}
t\hookrightarrow t/\lambda ^{2};  \TCItag{$1.4:3$}
\end{equation}
this is known as the van Hove or weak coupling limit in physics. One finds 
\begin{equation}
\lim_{\lambda \rightarrow 0}\langle \Psi _{R},\,B_{t/\lambda ^{2}}^{(\omega
,\lambda )}(g)\,B_{s/\lambda ^{2}}^{(\omega ,\lambda )\dagger }(f)\,\Psi
_{R}\rangle =\min (t,s)\int_{-\infty }^{\infty }d\tau \langle S_{\tau
}^{\omega }g,f\rangle .  \TCItag{$1.4:4$}
\end{equation}

Physically the limit $\lambda \to 0$ with $t \hookrightarrow t/\lambda^2 $
allows us to consider progressively weaker interactions which are allowed to
run over increasingly larger periods of time and so we obtain the long term
cumulative effect of the interaction on the system. Now the creation and
annihilation operators are gaussian in the vaccum state and, as a result, so
too are the operators $B^{(\omega ,\lambda ) \sharp }_{t/\lambda^2 }$.

Furthermore the limiting two-point function (29) is suggestive of the
correlation function of a Brownian motion. However, an interpretation of the
above in terms of classical Brownian motion is erroneous as it ignores the
essentially quantum probabilistic nature of these processes.

\subsection{\textbf{The Interaction}}

For technical reasons we work with a system Hamiltonian $H_{S}$ which has
discrete spectrum. $H_{R}^{1}$ is taken to be bounded below as required from
physics. The type of interaction $H_{I}$ which we wish to study is of the
form 
\begin{equation}
H_{I}=i\hbar \sum_{\omega \in F}\sum_{j=1}^{N(\omega )}\{\,D_{j}^{\omega
}\otimes A^{\dagger }(g_{j}^{\omega })-D_{j}^{\omega \dagger }\otimes
A(g_{j}^{\omega })\,\},  \TCItag{$1.5:1$}
\end{equation}
where $F$ is a discrete subset of $\mathbb{R}$. For each $\omega \in F$, we
take $D_{j}^{\omega }\in \mathcal{B}(\mathcal{H}_{S})$ to have harmonic free
evolution with frequency $\omega $: 
\begin{equation}
{\dfrac{1}{i\hbar }}[D_{j}^{\omega },H_{S}]=-i\omega D_{j}^{\omega
},\,j=1,...,N(\omega ).  \TCItag{$1.5:2$}
\end{equation}
Thus the superscript $\omega $ labels harmonic frequency and $%
j=1,...,N(\omega )$ the degeneracy of that frequency. An interaction similar
to (30) has been treated in [10], however there the test functions $%
g_{j}^{\omega }$ were taken to be equal for each value of $\omega $.

Our reasons for studying (30) above are because it allows us to treat the
most general interactions encountered in physics. Typically in quantum field
theory one considers an interaction of the type 
\begin{equation}
H_{I}=ih\int d^{3}k\,\{\theta (\mathbf{k})\otimes a^{\dagger }(\mathbf{k}%
)-\theta ^{\dagger }(\mathbf{k})\otimes a(\mathbf{k})\},  \TCItag{$1.5:3$}
\end{equation}
where $\{\theta (\mathbf{k}):\mathbf{k}\in \mathbb{R}^{3}\}$ is a family of
operators on $\mathcal{H}_{S}$. The operators $\theta (\mathbf{k})$ are
called the response terms: they contain local information about the
interaction. In the dipole approximation of quantum field theory one makes
the replacement 
\begin{equation}
\theta (\mathbf{k})\hookrightarrow \theta ^{\mathrm{dipole}}(\mathbf{k})=g(%
\mathbf{k})\theta (0),  \TCItag{$1.5:4$}
\end{equation}
where $g(\mathbf{k})$ is some suitable test function. The physical argument
is, cf. \cite{2}, that the response does not vary appreciably for values of
the wavelength of the reservoir particles which are large relative to the
physical dimensions of the system, though this can hardly be true for large
momenta. As a result one obtains the approximate Hamiltonian 
\begin{equation}
H_{I}\hookrightarrow H_{I}^{\mathrm{dipole}}=i\hbar \{\theta (0)\otimes
A^{\dagger }(g)-\theta ^{\dagger }(0)\otimes A(g)\}.  \TCItag{$1.5:5$}
\end{equation}
A further approximation often made by physicists is to replace $\theta (0)$
by an operator $D$ having a harmonic free-evolution with some frequency $%
\omega \in \mathbb{R}$. This approximation is just the rotating wave
approximation.

In order to avoid these appproximations we argue as follows: Let $B$ be a
complete basis of eigenstates of $H_{S}$, then 
\begin{equation}
H_{I}=\sum_{\phi ,\phi ^{\prime }\in B}\langle \phi |H_{I}|\phi ^{\prime
}\rangle |\phi \rangle \langle \phi ^{\prime }|,  \TCItag{$1.5:6$}
\end{equation}
but however we may write 
\begin{equation*}
\langle \phi |H_{I}|\phi ^{\prime }\rangle =i\hbar \int dk\{\langle \phi
|\theta (\mathbf{k})|\phi ^{\prime }\rangle a^{\dagger }(\mathbf{k})-\langle
\phi |\theta ^{\dagger }(\mathbf{k})|\phi ^{\prime }\rangle a(\mathbf{k})\}
\end{equation*}
\begin{equation}
\equiv i\hbar \lbrack A^{\dagger }(g_{\phi \phi ^{\prime }})-A(g_{\phi
^{\prime }\phi })],  \TCItag{$1.5:7$}
\end{equation}
where we have introduced the test functions 
\begin{equation}
g_{\phi \phi ^{\prime }}(\mathbf{k})=\langle \phi |\theta (\mathbf{k})|\phi
^{\prime }\rangle :  \TCItag{$1.5:8$}
\end{equation}
note that the order of $\phi $ and $\phi ^{\prime }$ is reversed in the
second term in due to the conjugate linear nature of the creation
feild.

This now means that the interaction can be expressed as 
\begin{equation}
H_{I}=i\hbar \sum_{\phi ,\phi ^{\prime }\in B}\{T_{\phi \phi ^{\prime
}}\otimes A^{\dagger }(g_{\phi \phi ^{\prime }})-T_{\phi \phi ^{\prime
}}^{\dagger }\otimes A(g_{\phi \phi ^{\prime }})\},  \TCItag{$1.5:9$}
\end{equation}
where we have introduced the transition operators $T_{\phi \phi ^{\prime
}}=|\phi \rangle \langle \phi ^{\prime }|$.

We note that the transition operators $T_{\phi \phi ^{\prime }}$ are
harmonic under the free evolution, in fact we have 
\begin{equation}
{\dfrac{1}{i\hbar }}[T_{\phi \phi ^{\prime }},H_{S}]=-i\omega _{\phi \phi
^{\prime }}T_{\phi \phi ^{\prime }},  \TCItag{$1.5:10$}
\end{equation}
where 
\begin{equation}
\omega _{\phi \phi ^{\prime }}={\dfrac{E_{\phi ^{\prime }}-E_{\phi }}{\hbar }%
}.  \TCItag{$1.5:11$}
\end{equation}
So $F=\{\omega _{\phi \phi ^{\prime }}:\phi ,\phi ^{\prime }\in B\}$ is now
the set of Bohr frequencies.

The expression (38) is now equivalent to the interaction (30) which we
propose to study. Here we need only relabel the $T_{\phi \phi ^{\prime }}$
as $D_{j}^{\omega }$ where $\omega =\omega _{\phi \phi ^{\prime }}$ and the $%
j$ again labels degeneracy. The functions $g_{\phi \phi ^{\prime }}$ are
relabeled accordingly.

\section{\textbf{THE QUANTUM STOCHASTIC LIMIT}}

\subsection{\textbf{\ Quantum Brownian Motions}}

In this section we first of all discuss the concept of quantum brownian
motion. As this is not yet widely known amongst physicists we give an
exposition below:

\textbf{Definition}\textit{.} \textit{A quantum brownian motion (QBM) is a
triple }$(H,\Phi ,(B_{t})_{t})$\textit{, where }$H$\textit{\ is a separable
Hilbert space, }$\Phi \in H$\textit{\ with }$\Vert \Phi \Vert =1$\textit{, }$%
(B_{t})_{t}$\textit{\ is a family of operators on }$H$\textit{\ such that}

\textit{(i) }$q_{t}=ReB_{t}$\textit{\ and }$p_{t}=ImB_{t}$\textit{\ are
classical brownian motions for the state }$\Phi $\textit{.}

\textit{(ii)}$[p_{s},q_{t}]=\dfrac{\kappa }{2i}\min (s,t),\,$\textit{\ where 
}$\kappa \in R$\textit{.}

The basic example is the following: Let $\mathcal{H}_{\mathbb{C}}=\Gamma
_{B}(L^{2}(\mathbb{R}))$ and $\Phi _{\mathbb{C}}$ be the vacuum state. Then
define $B_{t}$ to be 
\begin{equation}
B_{t}=A_{\mathbb{C}}(\chi _{\lbrack 0,t]})\   \TCItag{$2.1:1$}
\end{equation}
where $A_{\mathbb{C}}$ is the annihilation operator on $\mathcal{H}_{\mathbb{%
C}}$. From the (CCR) we have 
\begin{equation}
\lbrack B_{t},B_{s}^{\dagger }]=\langle \chi _{\lbrack 0,t]},\chi _{\lbrack
0,s]}\rangle =\min (t,s);\,[B_{t},B_{s}]=0=[B_{t}^{\dagger },B_{s}^{\dagger
}].  \TCItag{$2.1:2$}
\end{equation}
So setting $q_{t}={\dfrac{1}{2}}(B_{t}+B_{t}^{\dagger })$ and $p_{t}={\dfrac{%
1}{2i}}(B_{t}-B_{t}^{\dagger })$ we have from the (CCR) that $%
[p_{t},q_{s}]=-[q_{t},p_{s}]$ and 
\begin{equation}
\lbrack p_{t},q_{s}]={\dfrac{1}{2i}}\min (t,s).  \TCItag{$2.1:3$}
\end{equation}
Now if we set 
\begin{equation}
dB_{t}=B_{t+dt}-B_{t}=A_{\mathbb{C}}(\chi _{\lbrack t,t+dt]}),  \TCItag{$2.1:4$}
\end{equation}
we have that 
\begin{equation*}
\langle \Phi _{\mathbb{C}},dB_{t}^{\sharp }\Phi _{\mathbb{C}}\rangle
=0;\,\langle \Phi _{\mathbb{C}},(dB_{t}^{\sharp })^{2}\Phi _{\mathbb{C}%
}\rangle =0;\,\langle \Phi _{\mathbb{C}},dB_{t}^{\dagger }dB_{t}\Phi _{%
\mathbb{C}}\rangle =0;
\end{equation*}
\begin{equation}
\langle \Phi _{\mathbb{C}},dB_{t}dB_{t}^{\dagger }\Phi _{\mathbb{C}}\rangle
=dt.  \TCItag{$2.1:5$}
\end{equation}
Now $dB_{t}$ and $dB_{t}^{\dagger }$ are gaussian in the vacuum state,
because the creation and annihilation fields are, therefore so too are $%
dq_{t}$ and $dp_{t}$. Furthermore, 
\begin{equation}
\langle \Phi _{\mathbb{C}},(dq_{t})^{2}\Phi _{\mathbb{C}}\rangle ={\dfrac{1}{%
4}}\langle \Phi _{\mathbb{C}},(dB_{t}+dB_{t}^{\dagger })^{2}\Phi _{C}\rangle
={\dfrac{1}{4}}dt  \TCItag{$2.1:6$}
\end{equation}
and similarly $\langle \Phi _{\mathbb{C}},(dp_{t})^{2}\Phi _{C}\rangle ={%
\dfrac{1}{4}}dt$. Finally noting that at unequal times $s$ and $t$ 
\begin{equation}
\langle \Phi _{\mathbb{C}},dq_{t}dq_{s}\Phi _{\mathbb{C}}\rangle =0=\langle
\Phi _{\mathbb{C}},dp_{t}dp_{s}\Phi _{\mathbb{C}}\rangle ,  \TCItag{$2.1:7$}
\end{equation}
whenever $t\langle t+dt\leq s\langle s+ds$ or $s\langle s+ds\leq t\langle
t+dt$, we conclude that $(q_{t})_{t}$ and $(p_{t})_{t}$ are each separate
Brownian motions for expectations taken in the state $\Phi _{\mathbb{C}}$.
So $\{\mathcal{H}_{\mathbb{C}},\Phi _{\mathbb{C}},(B_{t})_{t}\}$ is a QBM.
We can introduce formal creation and annihilation densities $b^{\sharp }(t)$
satisfying 
\begin{equation}
\lbrack b(t),b(s)]=0=[b^{+}(t),b^{+}(s)];\,[b(t),b^{+}(s)]=\delta (t-s), 
\TCItag{$2.1:8$}
\end{equation}
such that 
\begin{equation}
A_{\mathbb{C}}(g)=\int_{\mathbb{R}}ds\,\overline{g(s)}b(s);\,A_{\mathbb{C}%
}^{\dagger }(g)=\int_{\mathbb{R}}ds\,g(s)b^{+}(s);  \TCItag{$2.1:9$}
\end{equation}
from this we see 
\begin{equation}
B_{t}^{\sharp }=\int_{0}^{t}ds\,b^{\sharp }(s).  \TCItag{$2.1:10$}
\end{equation}
We may write $b_{t}^{\sharp }={\dfrac{dB_{t}}{dt}}$ and consider these
densities as $``$quantum white noises''.

More generally let $K$ be a separable Hilbert space and let $L^{2}(\mathbb{R}%
,K)$ denote the set of square-integrable $K$-valued functions over $\mathbb{R%
}$. Now $h\in L^{2}(\mathbb{R},K)$ is a function $h(t)\in K$ with $\int_{%
\mathbb{R}}dt\,\Vert h(t)\Vert _{K}^{2}\langle \infty $. The inner product
on $L^{2}(\mathbb{R},K)$ is given by 
\begin{equation}
\langle h,h^{\prime }\rangle =\int_{\mathbb{R}}\langle h(t),h^{\prime
}(t)\rangle dt.  \TCItag{$2.1:11$}
\end{equation}
If $\{e_{n}\}_{n}$ is a complete orthonormal basis for $K$ then we can write 
$h(t)=\sum_{n}h_{n}(t)e_{n}$, where $h_{n}(t)=$ $\langle e_{n},h(t)\rangle
_{K}$; this gives a natural isomorphism 
\begin{equation}
L^{2}(\mathbb{R},K)\cong K\otimes L^{2}(\mathbb{R}).  \TCItag{$2.1:12$}
\end{equation}
Now take $\mathcal{H}_{K}=\Gamma _{B}(L^{2}(\mathbb{R},K))$ and let $\Phi
_{K}$ denote vacuum vector of $\mathcal{H}_{K}$, then a QBM is given by $(%
\mathcal{H}_{K},\Phi _{K},(B_{t}(g))_{t})$, for non-zero $g\in K$, where 
\begin{equation}
B_{t}(g)=A_{K}(g\otimes \chi _{\lbrack 0,t]})  \TCItag{$2.1:13$}
\end{equation}
where $A_{K}$ is the annihilation operator on $\mathcal{H}_{K}$. The
commutation relations are 
\begin{equation}
\lbrack B_{t}(g),B_{s}^{\dagger }(f)]=\langle g\otimes \chi _{\lbrack
0,t]},f\otimes \chi _{\lbrack 0,s]}\rangle =\langle g,f\rangle _{K}\min
(t,s),  \TCItag{$2.1:14$}
\end{equation}
with remaining commutators vanishing. So $\{\mathcal{H}_{K}=\Gamma
_{B}(L^{2}(\mathbb{R},K)),\,\Phi _{K},\,(B_{t}(g))_{t}\}$ is a QBM. Taking $%
K=\mathbb{C}$ and $|g|^{2}=\kappa $ leads back to the original example.

However, there is a more general possibility than that above; Let $Q\geq
1_{K}$ and set 
\begin{equation}
C=Q\otimes 1_{L^{2}(\mathbb{R})}.  \TCItag{$2.1:15$}
\end{equation}
Then let $\varphi _{C}$ be the state on the Weyl algebra $W(\mathcal{H}_{K})$
with covariance $C$. We can construct $\{G_{B}(\mathcal{H}_{K},C),$ $\pi _{%
\mathcal{H}_{K}}^{C},\Phi _{\mathcal{H}_{K}}^{C}\}$ the GNS triple over $\{W(%
\mathcal{H}_{K}),C\}$ and on it define the operator 
\begin{equation}
B_{Q}(g,t)=\pi _{\mathcal{H}_{K}}^{C}B_{t}(g).  \TCItag{$2.1:16$}
\end{equation}
Then $\{G_{B}(\mathcal{H}_{K},C),\Phi _{\mathcal{H}%
_{K}}^{C},(B_{Q}(g,t))_{t}\}$ is a QBM referred to as quantum Brownian
motion over $L^{2}(\mathbb{R},K)$ with covariance $C$, or more loosely with
covariance $Q$. We have that 
\begin{equation}
\langle \Phi _{\mathcal{H}_{K}}^{C},\,B_{Q}(g,t)\,B_{Q}^{\dagger
}(f,s)\,\Phi _{\mathcal{H}_{K}}^{C}\rangle =\varphi
_{C}(B_{t}(g)B_{s}^{\dagger }(f))  
=\min (t,s)\langle g,{\dfrac{Q+1}{2}}f\rangle _{K}, % \TCItag{$2.1:17$}
\end{equation}
and similarly 
\begin{equation}
\langle \Phi _{\mathcal{H}_{K}}^{C},\,B_{Q}^{\dagger
}(f,s)\,B_{Q}(g,t)\,\Phi _{\mathcal{H}_{K}}^{C}\rangle =\min (t,s)\langle g,{%
\dfrac{Q-1}{2}}f\rangle _{K}.  \TCItag{$2.1:18$}
\end{equation}
\textbf{\ }

\subsection{\textbf{Quantum Stochastic Calculus}}

As is well known, a stochastic calculus can be built up around classical
brownian motion and that the resulting theory has widespread applications to
studying noisy systems in physics and engineering. It is also possible to
build up a quantum stochastic calculus based on the QBMs we have just
considered. This was originally done by Hudson and Parthasarathy \cite{13}, 
\cite{14}. The basic integrators are $dt$ and, depending on the context, $%
dB_{t}^{\sharp }$ or $dB_{t}^{\sharp }(g)$ or $dB_{Q}^{\sharp }(t,g)$.

In the simplest case, for instance, we have, for a partition $-\infty
=t_{1}<t_{2}<...<t_{n}<t_{n+1}=\infty $, 
\begin{equation*}
L^{2}(\mathbb{R})=\bigoplus_{m=1}^{n}L^{2}([t_{m},t_{m+1}])
\end{equation*}
and consequently 
\begin{equation}
\Gamma _{B}(L^{2}(\mathbb{R}))=\bigotimes_{m=1}^{n}\Gamma
_{B}(L^{2}([t_{m},t_{m+1}]).  \TCItag{$2.2:1$}
\end{equation}
This gives the required time filtration in the quantum situation. We say
that a family of operators $(X_{t})_{t}$ on $\Gamma _{B}((L^{2}(\mathbb{R})))
$ is adapted if, for all $t$, 
\begin{equation}
X_{t}\equiv {\tilde{X}}_{t}\otimes 1  \TCItag{$2.2:2$}
\end{equation}
on $\Gamma _{B}(L^{2}((-\infty ,t))\otimes \Gamma _{B}(L^{2}([t,\infty )))$.
The quantum Ito table reads as 
\begin{eqnarray*}
dB_{Q}(g,t)\,dB_{Q}^{\dagger }(f,t) &\equiv &\langle g,{\dfrac{Q+1}{2}}%
f\rangle _{K}dt, \\
dB_{Q}^{\dagger }(f,t)\,dB_{Q}(g,t) &\equiv &\langle g,{\dfrac{Q-1}{2}}%
f\rangle _{K}dt,
\end{eqnarray*}
\begin{equation}
(dt)^{2},\,dt\,dB_{Q}^{\sharp }(g,t),\,(dB_{Q}^{\sharp }(g,t))^{2}\equiv 0, 
\TCItag{$2.2:3$}
\end{equation}
Let $(X_{t})_{t}$ be an adapted process and of the form 
\begin{equation}
X_{t}=\int_{0}^{t}(x_{s}ds+x_{s}^{+}dB_{Q}^{\dagger
}(g,s)+x_{s}^{-}dB_{Q}(g,s))  \TCItag{$2.2:4$}
\end{equation}
and $(Y_{t})_{t}$ a similar process, then we have the quantum Ito formula 
\begin{equation}
d(X_{t}.Y_{t})\equiv dX_{t}.Y_{t}+X_{t}.dY_{t}+dX_{t}.dY_{t}  \TCItag{$2.2:5$}
\end{equation}
with 
\begin{equation}
dX_{t}=x_{t}dt+x_{t}^{+}dB_{Q}^{\dagger }(g,t)+x_{t}^{-}dB_{Q}(g,t)). 
\TCItag{$2.2:6$}
\end{equation}

\subsection{\textbf{The Weak Coupling Limit of Quantum Field Theory}}

The results of Accardi, Frigerio and Lu concerning the weak coupling limit
for an interaction (31) which has undergone both a dipole and rotating
wave approximation can be summarised as follows;

Recall that 
\begin{equation*}
\lim_{\lambda \rightarrow 0}\langle \Psi _{R},B_{t/\lambda ^{2}}^{(\omega
,\lambda )}(g)B_{s/\lambda ^{2}}^{(\omega ,\lambda )\dagger }(f)\Psi
_{R}\rangle =\min (t,s)\int_{-\infty }^{\infty }d\tau \langle S_{\tau
}^{\omega }g,f\rangle ;
\end{equation*}
now define a sesquilinear form $(.|.)^{\omega }$ on $\mathcal{H}_{R}^{1}$
the one particle reservoir space by 
\begin{equation}
(g|f)^{\omega }=\int_{-\infty }^{\infty }d\tau \langle S_{\tau }^{\omega
}g,f\rangle .  \TCItag{$2.3:1$}
\end{equation}
We consider the space of suitable test-functions $T^{\omega }\subset 
\mathcal{H}_{R}^{1}$, determined by the condition 
\begin{equation}
\int_{-\infty }^{\infty }dt\,|\langle g,S_{t}^{\omega }f\rangle |\langle
\infty ,  \TCItag{$2.3:2$}
\end{equation}
whenever $f,g\in T^{\omega }$. Note that technically $T^{\omega }$ does not
depend on $\omega $ however we keep it in as a label. Then we construct $%
K_{\omega }$ the completion of $T^{\omega }$ with respect to $(.|.)^{\omega }
$. That is $K^{\omega }$ is the completion of $T^{\omega }$ factored out by
its $(.|.)^{\omega }$-norm null space. $K^{\omega }$ is a separable Hilbert
space with inner product $(.|.)^{\omega }$

\textbf{Theorem 1.} I\textit{n the limit }$\lambda \rightarrow 0$\textit{\
the stochastic process on the resevoir space } 
\begin{equation*}
\{\mathcal{H}_{R},\Psi _{R},(B_{t/\lambda ^{2}}^{(\omega ,\lambda
)}(f))_{t}\},
\end{equation*}
\textit{for }$f\in K^{\omega }$\textit{, converges weakly in the sense of
matrix elements to QBM on }$L^{2}(R,K_{\omega })$\textit{. We denote this
QBM by }$\{H^{\omega }=\Gamma _{B}(L^{2}(R,K^{\omega })),\Phi ^{\omega
}=\Phi _{K^{\omega }},(B_{t}^{\omega }(f))_{t}\}$\textit{\ In the next
theorem we show that }$U_{t/\lambda ^{2}}^{(\lambda )}$\textit{\ converges
to a stochastic process }$U_{t}$\textit{\ on }$H_{S}\otimes H^{\omega }$%
\textit{\ in a sense to be made explicit now.}

\bigskip

\textbf{Theorem 2.} \textit{Let }$f^{(j)},h^{(j^{\prime })}\in K^{\omega }$%
\textit{; }$T^{(j)},S^{(j^{\prime })}\rangle 0$\textit{, for }$%
j=1,...,n:j^{\prime }=1,...,m$\textit{\ and let }$\phi ,\phi ^{\prime }\in
H_{S}$\textit{\ then the limit as }$\lambda \rightarrow 0$\textit{\ of the
matrix element } 
\begin{equation}
\langle \phi \otimes B_{T^{(1)}/\lambda ^{2}}^{(\omega ,\lambda )\dagger
}(f^{(1)})...B_{T^{(m)}/\lambda ^{2}}^{(\omega ,\lambda )\dagger
}(f^{(n)})\Psi _{R}|  U_{t/\lambda ^{2}}^{(\lambda )}|\phi ^{\prime }\otimes B_{S^{(1)}/\lambda
^{2}}^{(\omega ,\lambda )\dagger }(h^{(1)})...B_{S^{(m)}/\lambda
^{2}}^{(\omega ,\lambda )\dagger }(h^{(m)})\Psi _{R}\rangle  
%\TCItag{$2.3:3$}
\end{equation}
\textit{exists and equals } 
\begin{equation}
\langle \phi \otimes B_{T^{(1)}}^{\omega \dagger
}(f^{(1)})...B_{T^{(n)}}^{\omega \dagger }(f^{(n)})\Phi ^{\omega
}|U_{t}|\phi ^{\prime }\otimes B_{S^{(1)}}^{\omega \dagger
}(h^{(1)})...B_{S^{(m)}}^{\omega \dagger }(h^{(m)})\Phi ^{\omega }\rangle , 
\TCItag{$2.3:4$}
\end{equation}
\textit{where }$U_{t}$\textit{\ is a process on }$H_{S}\otimes H^{\omega }$%
\textit{\ which is the solution to the quantum stochastic differential
equation } 
\begin{equation}
dU_{t}=\{D\otimes dB_{t}^{\omega \dagger }(g)-D^{\dagger }\otimes
dB_{t}^{\omega }(g)-(g|g)^{\omega -}D^{\dagger }D\otimes dt\}U_{t}, 
\TCItag{$2.3:5$}
\end{equation}
\textit{with } 
\begin{equation}
(g|f)^{\omega -}=\int_{-\infty }^{0}d\tau \langle g,S_{\tau }^{\omega
}f\rangle .  \TCItag{$2.3:6$}
\end{equation}
Note that $d(U_{t}U_{t}^{\dagger })\equiv 0\equiv d(U_{t}^{\dagger }U_{t})$
by the quantum Ito formula and the Ito table. So $U_{t}$ is \textit{unitary}
on $\mathcal{H}_{S}\otimes \mathcal{H}^{\omega }$, however it describes an 
\textit{irreversible} evolution when restricted to $\mathcal{H}_{S}$. The
unitarity condition corresponds to a fluctuation-dissipation law.

\textbf{Theorem 3} \textit{Let }$X\in B(H_{S})$\textit{, then in the
notation of theorem 2 the limit } 
\begin{equation*}
\langle \phi \otimes B_{T^{(1)}/\lambda ^{2}}^{(\omega ,\lambda )\dagger
}(f^{(1)})..B_{T^{(n)}/\lambda ^{2}}^{(\omega ,\lambda )\dagger
}(f^{(n)})\Psi _{R}|
\end{equation*}
\begin{equation}
U_{t/\lambda ^{2}}^{(\lambda )\ \dagger }\ (X\otimes 1_{R})U_{t/\lambda
^{2}}^{(\lambda )}|\phi ^{\prime }\otimes B_{S^{(1)}/\lambda ^{2}}^{(\omega
,\lambda )\dagger }(h^{(1)})...B_{S^{(m)}/\lambda ^{2}}^{(\omega ,\lambda
)\dagger }(h^{(m)})\Psi _{R}\rangle  \TCItag{$2.3:7$}
\end{equation}
\textit{exists and equals } 
\begin{equation}
\langle \phi \otimes B_{T^{(1)}}^{\omega \dagger
}(f^{(1)})...B_{T^{(n)}}^{\omega \dagger }(f^{(n)})\Phi ^{\omega
}|U_{t}^{\dagger }(X\otimes 1)U_{t}|\phi ^{\prime }\otimes B_{(1)}^{\omega
\dagger }(h^{(1)})...B_{(m)}^{\omega \dagger }(h^{(m)})\Phi ^{\omega
}\rangle .  \TCItag{$2.3:8$}
\end{equation}
Note that in these theorems we encounter vectors of the type $B_{T/\lambda
^{2}}^{(\omega ,\lambda )\dagger }(f)\Psi _{R}$ which are exponential
vectors with test functions $\lambda \int_{0}^{T/\lambda ^{2}}d\tau S_{\tau
}^{\omega }f$: they are called $\mathit{collective\ coherent\ vectors}$ in
the terminology of Accardi and Lu and they are designed to extract the long
time cumulative behaviour of the reservoir fields.

\subsection{\textbf{Non-Zero Temperature Reservoir}}

Next, for the non-vacuum case, we consider a density matrix $\rho _{Q}$ on $%
\mathcal{H_{R}}$ which is invariant under the free evolution and gaussian
with covariance $Q\geq 1_{\mathcal{H}_{R}^{1}}$. That is 
\begin{equation}
\mathrm{Tr}\{\rho _{Q}W(g)\}=e^{-{\dfrac{1}{2}}\langle g,Qg\rangle
},\,\forall g\in \mathcal{H}_{R}^{1}.  \TCItag{$2.4:1$}
\end{equation}
The invariance condition is equivalent to 
\begin{equation}
\lbrack S_{t},Q]=0,\mathrm{on\,Dom}(Q).  \TCItag{$2.4:2$}
\end{equation}
In particular, the choice of a heat bath at inverse temperature $\beta $ and
fugacity $z$ is given by 
\begin{equation}
Q={\dfrac{1+ze^{-\beta H_{R}^{1}}}{1-ze^{-\beta H_{R}^{1}}}},  \TCItag{$2.4:3$}
\end{equation}
that is 
\begin{equation}
(Qf)(k)=\mathrm{coth}{\dfrac{\beta }{2}}(\hbar \omega (k)-\mu )\,f(k), 
\TCItag{$2.4:4$}
\end{equation}
where $\mu ={\dfrac{1}{\beta }}\ln z$ is the chemical potential.

Now 
\begin{equation*}
\lim_{\lambda \rightarrow 0}\mathrm{Tr}\{\rho _{Q}\,B_{t/\lambda
^{2}}^{(\omega ,\lambda )}(g)B_{s/\lambda ^{2}}^{(\omega ,\lambda )\dagger
}(f)\}=\min (t,s)\int_{-\infty }^{\infty }d\tau \mathrm{Tr}\{\rho
_{Q}\,A(S_{\tau }^{\omega }g)A^{\dagger }(f)\}
\end{equation*}
\begin{equation}
=\min (t,s)\int_{-\infty }^{\infty }d\tau \langle S_{\tau }^{\omega }g,({%
\dfrac{Q+1}{2}})f\rangle ,  \TCItag{$2.4:5$}
\end{equation}
and similarly 
\begin{equation}
\lim_{\lambda \rightarrow 0}\mathrm{Tr}\{\rho _{Q}\,B_{s/\lambda
^{2}}^{(\omega \lambda )\dagger }(f)B_{t/\lambda ^{2}}^{(\omega ,\lambda
)}(g)\}=\min (t,s)\int_{-\infty }^{\infty }d\tau \langle S_{\tau }^{\omega
}g,({\dfrac{Q-1}{2}})f\rangle .  \TCItag{$2.4:6$}
\end{equation}
Let $T_{Q}^{\omega }$ be the subset of Dom$(Q)$ such that 
\begin{equation}
\int_{-\infty }^{\infty }|\langle f,S_{t}^{\omega }h\rangle |dt <
\infty \,\mathrm{and}\,\int_{-\infty }^{\infty }|\langle f,S_{t}^{\omega
}Qh\rangle |dt < \infty ,  \TCItag{$2.4:7$}
\end{equation}
whenever $f,h\in T_{Q}^{\omega }$. Let $K_{Q}^{\omega }$ be the Hilbert
space completion of $T_{Q}^{\omega }$ with respect to the sesquilinear form $%
(.|.)_{Q}^{\omega }$ given by 
\begin{equation}
(f|h)_{Q}^{\omega }=\int_{-\infty }^{\infty }\langle f,QS_{t}^{\omega
}h\rangle dt.  \TCItag{$2.4:8$}
\end{equation}
Note that in most cases $T_{Q}^{\omega }$ is dense in $\mathcal{H}_{R}^{1}$
and that $K_{Q}^{\omega }$ is a Hilbert space equipped with inner product $%
(.|.)_{Q}^{\omega }$.

\textbf{Theorem 1a}: \textit{The process }$(B_{t/\lambda ^{2}}^{(\lambda
)}(f))_{t}$\textit{\ in the mixed state }$\rho _{Q}$\textit{\ converges
weakly in the sense of matrix elements to QBM over }$L^{2}(R,K_{Q}^{\omega })
$\textit{\ with covariance }$Q$\textit{. This is denoted as }$%
\{H_{Q}^{\omega }=G_{B}((L^{2}(R,K_{Q}^{\omega }),Q\otimes 1),\,\Phi
_{Q}^{\omega },\,(B_{Q}^{\omega }(f,t))_{t}\}$\textit{. }

\bigskip 

\textbf{Theorem 2a}; \textit{For }$\phi ,\phi ^{\prime }\in
H_{S},\,f^{(j)},h^{(j^{\prime })}\in K_{Q}^{\omega }$\textit{\ and }$%
T^{(j)},S^{(j^{\prime })}\rangle 0$\textit{, for }$j=1,...n;j^{\prime
}=1,...,m$\textit{, the limit as }$\lambda \rightarrow 0$\textit{\ of } 
\begin{equation}
\mathrm{Tr}\{\,|\phi ^{\prime }\rangle \langle \phi |\otimes
B_{S^{(m)}/\lambda ^{2}}^{(\omega ,\lambda )\dagger
}(h^{(m)})...B_{S^{(1)}/\lambda ^{2}}^{(\omega ,\lambda )\dagger
}(h^{(1)})\rho _{Q}B_{T^{(1)}/\lambda ^{2}}^{(\omega ,\lambda
)}(f^{(1)})...B_{T^{(n)}/\lambda ^{2}}^{(\omega ,\lambda
)}(f^{(n)})\,U_{t/\lambda ^{2}}^{(\lambda )}\}  \TCItag{$2.4:9$}
\end{equation}
\textit{exists and equals } 
\begin{equation}
\langle \phi \otimes B_{Q}^{\omega \dagger
}(f^{(1)},T^{(1)})...B_{Q}^{\omega \dagger }(f^{(n)},T^{(n)})\Phi
_{Q}^{\omega }|U_{t}|\phi ^{\prime }\otimes B_{Q}^{\omega \dagger
}(h^{(1)},S^{(1)})...B_{Q}^{\omega \dagger }(h^{(m)},S^{(m)})\Phi
_{Q}^{\omega }\rangle   \TCItag{$2.4:10$}
\end{equation}
\textit{where }$U_{t}$\textit{\ is a unitary operator on }$H_{S}\otimes
G_{B}(L^{2}(R,K_{Q}^{\omega }),C)$\textit{, with }$C=Q\otimes 1$\textit{,
satisfying the QSDE } 
\begin{equation}
dU_{t}=\{D\otimes dB_{Q}^{\omega \dagger }(g,t)-D^{\dagger }\otimes
dB_{Q}^{\omega }(g,t)-(g|g)_{Q+}^{\omega -}D^{\dagger }D\otimes dt-\overline{%
(g|g)}_{Q-}^{\omega -}DD^{\dagger }\otimes dt\}U_{t},\   \TCItag{$2.4:11$}
\end{equation}
\textit{where } 
\begin{equation}
(g|f)_{Q\pm }^{\omega -}=\int_{-\infty }^{0}d\tau \langle g,S_{\tau
}^{\omega }{\dfrac{Q\pm 1}{2}}f\rangle .  \TCItag{$2.4:12$}
\end{equation}
\textbf{Theorem 3a}:\textit{\ For }$X\in B(H_{S})$\textit{, then in the
notation of theorem 2} 
\begin{equation}
\lim_{\lambda \rightarrow 0}\mathrm{Tr}\{|\phi ^{\prime }\rangle \langle
\phi |\otimes B_{S^{(m)}/\lambda ^{2}}^{(\omega ,\lambda )\dagger
}(h^{(m)})...B_{S^{(1)}/\lambda ^{2}}^{(\omega ,\lambda )\dagger
}(h^{(1)})\rho _{Q}
B_{T^{(1)}/\lambda ^{2}}^{(\omega ,\lambda )}(f^{(1)})...B_{T^{(n)}/\lambda
^{2}}^{(\omega ,\lambda )}(f^{(n)})\,u_{t/\lambda ^{2}}^{(\lambda
)}(X\otimes 1)\}  \TCItag{$2.4:13$}
\end{equation}
\textit{exists and equals} 
\begin{equation}
\langle \phi \otimes B_{Q}^{\omega \dagger
}(f^{(1)},T^{(1)})...B_{Q}^{\omega \dagger }(f^{(n)},T^{(n)})\Phi
_{Q}^{\omega }|   
U_{t}^{\dagger }(X\otimes 1)U_{t}|\phi ^{\prime }\otimes B_{Q}^{\omega
\dagger }(h^{(1)},S^{(1)})...B_{Q}^{\omega \dagger }(h^{(m)},S^{(m)})\Phi
_{Q}^{\omega }\rangle .  \TCItag{$2.4:14$}
\end{equation}

\bigskip

\subsection{\textbf{The Quantum Stochastic Limit for the Full Interaction}}

Now suppose that the interaction is of the form (31). The problem of
dropping the rotating wave approximation was first tackled by Accardi and Lu
in \cite{10} for an interaction similar to (31) except that all the test
functions were taken to be the same. The result is that for each Bohr
frequency $\omega $ we obtain a separate independent QBM.

First of all note that $\Gamma _{B}(\bigoplus_{\omega \in F}L^{2}(\mathbb{R}%
,K_{Q}^{\omega }))=\bigotimes_{\omega \in F}\Gamma _{B}(L^{2}(\mathbb{R}%
,K_{Q}^{\omega }))$ then consider the Weyl algebra $W(\bigoplus_{\omega \in
F}L^{2}(\mathbb{R},K_{Q}^{\omega }))=\bigotimes_{\omega \in F}W(L^{2}(%
\mathbb{R},K_{Q}^{\omega })),$ with quasi-free state $\varphi _{\tilde{C}}$
with covariance ${\tilde{C}}=\bigotimes_{\omega \in F}C$, where $C=Q\otimes
1 $ on $\mathcal{H}_{K_{Q}^{\omega }}\cong K_{Q}^{\omega }\otimes L^{2}(%
\mathbb{R})$.

The GNS triple over $\{W(\bigoplus_{\omega \in F}L^{2}(\mathbb{R}%
,K_{Q}^{\omega })),\varphi _{\tilde{C}}\}$ is $\{\mathcal{H}_{Q}^{\omega }=$ 
$\ \ \Gamma _{B}(\bigoplus_{\omega \in F}L^{2}(\mathbb{R},K_{Q}^{\omega })),{%
\tilde{C}})),\pi _{Q}^{F},$ $\Phi _{Q}^{\omega }\}$. Now observe that 
\begin{equation}
\mathcal{H}_{Q}^{F}=\bigotimes_{\omega \in F}\mathcal{H}_{Q}^{\omega },\,\pi
_{Q}^{F}=\bigotimes_{\omega \in F}\pi _{Q}^{\omega };\,\Phi
_{Q}^{F}=\bigotimes_{\omega \in F}\Phi _{Q}^{\omega }.  \TCItag{$2.5:1$}
\end{equation}
For each $f\in K_{Q}^{\omega }$ we have 
\begin{equation}
B_{Q}^{\omega }(f,t)=\pi _{Q}^{\omega }A_{Q}^{\omega }(f\otimes \chi
_{\lbrack 0,t]}),  \TCItag{$2.5:2$}
\end{equation}
where $A_{Q}^{\omega }$ is annihilation operator on $L^{2}(\mathbb{R},{\
K_{Q}^{\omega }}),$ so for $t_{\omega }\rangle 0,f_{\omega }\in
K_{Q}^{\omega }$, for each $\omega \in F$, we have 
\begin{equation}
B_{Q}^{F}(\bigotimes_{\omega \in F}f_{\omega },(t_{\omega })_{\omega \in
F})=\bigotimes_{\omega \in F}B_{Q}^{\omega }(f_{\omega },t_{\omega }). 
\TCItag{$2.5:3$}
\end{equation}

\bigskip 

\textbf{Theorem 1b}: \textit{For each }$\omega \in F$\textit{\ and }$f\in
K_{Q}^{\omega }$\textit{\ the limit }$\lambda \rightarrow 0$\textit{, }$%
B_{t/\lambda ^{2}}^{(\omega ,\lambda )}(f)$\textit{\ taken in the state }$%
\rho _{Q}$\textit{\ converges in the sense of matrix elements to QBM over }$%
L^{2}(R,K_{Q}^{\omega })$\textit{\ with covariance }$Q$\textit{\ and each of
these limiting processes are independent for different values of }$\omega $%
\textit{.}

\textbf{Theorem 2b}: \textit{Let }$f_{\omega }^{(j)},h_{\omega }^{(j^{\prime
})}\in K_{Q}^{\omega };T_{\omega }^{(j)},S_{\omega }^{(j^{\prime })}\rangle 0
$\textit{\ for each }$\omega \in F\,j=1,...,n_{\omega };j^{\prime
}=1,...,m_{\omega };\,and\ t\geq 0;\ \phi ,\phi ^{\prime }\in H_{S}$\textit{%
\ then the limit as }$\lambda \rightarrow 0$\textit{\ of the matrix element} 
\begin{equation}
\mathrm{Tr}\{|\phi ^{\prime }\rangle \langle \phi |\otimes \lbrack
\bigotimes_{\omega \in F}B_{S_{\omega }^{(m_{\omega })}/\lambda
^{2}}^{(\omega ,\lambda )}(h_{\omega }^{(m_{\omega })})...B_{S_{\omega
}^{(1)}/\lambda ^{2}}^{(\omega ,\lambda )}(h_{\omega }^{(1)})]^{\dagger } 
\times \rho _{Q}[\bigotimes_{\omega \in F}B_{T_{\omega }^{(1)}/\lambda
^{2}}^{(\omega ,\lambda )}(f_{\omega }^{(1)})...B_{T_{\omega }^{(n_{\omega
})}/\lambda ^{2}}^{(\omega ,\lambda )}(f_{\omega }^{(n_{\omega
})})]\,U_{t/\lambda ^{2}}^{(\lambda )}\}  \TCItag{$2.5:4$}
\end{equation}
\textit{exists and equals} 
\begin{equation}
\langle \phi \otimes \lbrack \bigotimes_{\omega \in F}B_{Q}^{\omega
}(f_{\omega }^{(1)},T_{\omega }^{(1)})...B_{Q}^{\omega }(f_{\omega
}^{(n_{\omega })},T_{\omega }^{(n_{\omega })})]^{\dagger }\Phi _{Q}^{F}|
\end{equation}
\begin{equation}
U_{t}|\phi ^{\prime }\otimes \lbrack \bigotimes_{\omega \in F}B_{Q}^{\omega
}(h_{\omega }^{(1)},S_{\omega }^{(1)})...B_{Q}^{\omega }(h_{\omega
}^{(m_{\omega })},S_{\omega }^{(m_{\omega })})]^{\dagger }\Phi
_{Q}^{F}\rangle ,  \TCItag{$2.5:5$}
\end{equation}
\textit{where }$U_{t}$\textit{\ is unitary on }$H_{S}\otimes H_{Q}^{F}$%
\textit{\ and satisfies quantum stochastic differential equation } 
\begin{equation*}
dU_{t}=\{\sum_{\omega \in F}\sum_{j=1}^{N(\omega )}\ [D_{j}^{\omega }\otimes
dB_{\omega }^{\dagger }(g_{j}^{\omega },t)-D_{j}^{\omega \dagger }\otimes
dB_{\omega }(g_{j}^{\omega },t)]
\end{equation*}
\begin{equation}
-\sum_{\omega \in F}\ \sum_{j,k=1}^{N(\omega )}\ D_{j}^{\omega \dagger
}D_{k}^{\omega }\ (g_{j}^{\omega }|g_{k}^{\omega })_{Q+}^{\omega -}\
dt-\sum_{\omega \in F}\ \sum_{j,k=1}^{N(\omega )}\ D_{j}^{\omega
}D_{k}^{\omega \dagger }\ \overline{(g_{j}^{\omega }|g_{k}^{\omega })}%
_{Q-}^{\omega -}\ dt\}\ U_{t},  \TCItag{$2.5:6$}
\end{equation}
\textit{with }$U_{0}=1$\textit{. The coefficients are given by } 
\begin{equation}
(f|h)_{Q\pm }^{\omega -}=\int_{-\infty }^{0}dt\langle f,S_{t}^{\omega }({%
\dfrac{Q\pm 1}{2}})h\rangle ,  \TCItag{$2.5:7$}
\end{equation}
\textit{for }$f,h\in K_{\omega }$\textit{. The Ito table is given by } 
\begin{equation*}
dB_{Q}^{\omega }(f,t)dB_{Q}^{\omega ^{\prime }\dagger }(g,t)\equiv \delta
_{\omega ,\omega ^{\prime }}(f|g)_{Q+}^{\omega }dt
\end{equation*}
\begin{equation}
dB_{Q}^{\omega \dagger }(g,t)dB_{Q}^{\omega ^{\prime }}(f,t)\equiv \delta
_{\omega ,\omega ^{\prime }}(f|g)_{Q-}^{\omega }dt.  \TCItag{$2.5:8$}
\end{equation}
\textit{The coefficients in (2.5:8) are } 
\begin{equation*}
(f|g)_{Q+}^{\omega }=\int_{-\infty }^{\infty }dt\langle f,S_{t}^{\omega }({%
\dfrac{Q+1}{2}})g\rangle =(f|g)_{Q+}^{\omega -}+\overline{(g|f)_{Q+}^{\omega
-}};
\end{equation*}
\begin{equation}
(f|g)_{Q-}^{\omega }=\int_{-\infty }^{\infty }dt\langle f,S_{t}^{\omega }({%
\dfrac{Q-1}{2}})g\rangle =\overline{(g|f)_{Q-}^{\omega -}}%
+(f|g)_{Q-}^{\omega -}.  \TCItag{$2.5:9$}
\end{equation}

The proof of theorem 2a is as follows; first of all we know that for
different $\omega $ we can always consider independent Q-quantum Brownian
motions. This is done in \cite{10}. The next step is to consider the effect
of the degeneracy which may arise for each $\omega \in F$. In this case we
must, therefore, generalize the results of \cite{8},\cite{10} accordingly.
This is done in appendices B and C.

\bigskip 

\textbf{Theorem 3a.} \textit{In the notations of theorem 2a, for any }$X\in
B(H_{S})$\textit{, the limit as }$\lambda \rightarrow 0$\textit{\ } 
\begin{equation}
\mathrm{Tr}\{|\phi ^{\prime }\rangle \langle \phi |\otimes \lbrack
\bigotimes_{\omega \in F}B_{S_{\omega }^{(1)}/\lambda ^{2}}^{(\omega
,\lambda )}(h_{\omega }^{(1)})...B_{S_{\omega }^{(m_{\omega })}/\lambda
^{2}}^{(\omega ,\lambda )}(h_{\omega }^{(m_{\omega })})]^{\dagger }\ 
\rho _{Q}[\bigotimes_{\omega \in F}B_{T_{\omega }^{(1)}/\lambda
^{2}}^{(\omega ,\lambda )}(f_{\omega }^{(1)})...B_{T_{\omega }^{(n_{\omega
})}/\lambda ^{2}}^{(\omega ,\lambda )}(f_{\omega }^{(n_{\omega })})]\text{ }%
u_{t/\lambda ^{2}}^{(\lambda )}(X\otimes 1)\}  \TCItag{$2.5:10$}
\end{equation}
\textit{exists and equals } 
\begin{equation}
\langle \phi \otimes [ \bigotimes_{\omega \in F}
B_{Q}^{\omega} 
(f_{\omega }^{(1)},T_{\omega }^{(1)})
...B_{Q}^{\omega }(f_{\omega}^{(n_{\omega })}
,T_{\omega }^{(n_{\omega })})]^{\dagger }\Phi _{Q}^{F}|
U_{t}^{\dagger }(X\otimes 1)U_{t}|\phi ^{\prime }\otimes [
\bigotimes_{\omega \in F}B_{Q}^{\omega }(h_{\omega }^{(1)},S_{\omega
}^{(1)})...B_{Q}^{\omega }(h_{\omega }^{(m_{\omega })},S_{\omega
}^{(m_{\omega })})]^{\dagger }\Phi _{Q}^{F} \rangle ,  \TCItag{$2.5:11$}
\end{equation}
\textit{where }$U_{t}$\textit{\ is the solution to the quantum stochastic
differential equation of theorem.}

\subsection{\textbf{The Langevin and Master Equations}}

In each of the cases the right hand side of the
expression for $dU_{t}$ contains a term of the form $-Y\otimes dtU_{t}$. For
instance,  we have 
\begin{equation}
Y=\sum_{\omega \in F}\ \sum_{j,k=1}^{N(\omega )}\{\ D_{j}^{\omega \dagger
}D_{k}^{\omega }\ (g_{j}^{\omega }|g_{k}^{\omega })_{Q+}^{\omega -}\
+D_{j}^{\omega }D_{k}^{\omega \dagger }\ \overline{(g_{j}^{\omega
}|g_{k}^{\omega })}_{Q-}^{\omega -}\ \},  \TCItag{$2.6:1$}
\end{equation}
The Langevin equation then reads as follows 
\begin{eqnarray*}
d[U_{t}^{\dagger }(X\otimes 1)U_{t}] &\equiv &[dU_{t}]^{\dagger }(X\otimes
1)U_{t}+U_{t}^{\dagger }(X\otimes 1)dU_{t}+[dU_{t}]^{\dagger }(X\otimes
1)dU_{t} \\
&\equiv &U_{t}^{\dagger }[L_{0}(X)\otimes dt+\sum_{\omega \in F}\
\sum_{j,k=1}^{N(\omega )}L_{j+}^{\omega }(X)\otimes dB_{Q}^{\omega \dagger
}(g_{j}^{\omega },t)
\end{eqnarray*}
\begin{equation}
\ +\sum_{\omega \in F}\ \sum_{j,k=1}^{N(\omega )}L_{j-}^{\omega }(X)\otimes
dB_{Q}^{\omega }(g_{j}^{\omega },t)\}U_{t},  \TCItag{$2.6:2$}
\end{equation}
where 
\begin{equation}
L_{0}(X)=-XY-Y^{\dagger }X+\Theta (X),  \TCItag{$2.6:3$}
\end{equation}
with 
\begin{eqnarray*}
\Theta (X) &=&\sum_{\omega \in F}\ \sum_{j,k=1}^{N(\omega )}\{\
D_{j}^{\omega \dagger }XD_{k}^{\omega }\ [(g_{j}^{\omega }|g_{k}^{\omega
})_{Q+}^{\omega -}\ +\overline{(g_{k}^{\omega }|g_{j}^{\omega })}%
_{Q+}^{\omega -}] +D_{j}^{\omega }XD_{k}^{\omega \dagger }\ [\overline{(g_{j}^{\omega
}|g_{k}^{\omega })}_{Q-}^{\omega -}+(g_{k}^{\omega }|g_{j}^{\omega
})_{Q-}^{\omega -}]\ \}
\end{eqnarray*}
\begin{equation}
=\sum_{\omega \in F}\ \sum_{j,k=1}^{N(\omega )}\{\ D_{j}^{\omega \dagger
}XD_{k}^{\omega }\ (g_{j}^{\omega }|g_{k}^{\omega })_{Q+}^{\omega }\
+D_{k}^{\omega }XD_{j}^{\omega \dagger }\ (g_{j}^{\omega }|g_{k}^{\omega
})_{Q-}^{\omega }]\ \},  \TCItag{$2.6:4$}
\end{equation}
and 
\begin{equation}
L_{j+}^{\omega }(X)=XD_{j}^{\omega }-D_{j}^{\omega }X,\,L_{j-}^{\omega
}(X)=D_{j}^{\omega \dagger }X-XD_{j}^{\omega \dagger }.  \TCItag{$2.6:5$}
\end{equation}
Note that unitarity follows from 
\begin{equation}
L_{0}(1_{S})=-(Y+Y^{\dagger })+\Theta (1_{S})=0;\ L_{j\pm }^{\omega
}(1_{S})=0.  \TCItag{$2.6:6$}
\end{equation}
It is instructive to set $Y={\dfrac{1}{2}}\Gamma +{\dfrac{i}{\hbar }}%
H_{S}^{\prime }$ where both $\Gamma $ and $H_{S}^{\prime }$ are
self-adjoint; we then have that 
\begin{equation}
L_{0}(X)=-{\dfrac{1}{2}}(X\Gamma +\Gamma X)+\Theta (X)+{\dfrac{1}{i\hbar }}%
[X,H_{S}^{\prime }].  \TCItag{$2.6:7$}
\end{equation}
The unitarity condition is then $2\mathrm{Re}\ Y=\Gamma =\Theta (1_{S})$;
this is the fluctuation-dissipation relation. The presence of the imaginary
term $H_{S}^{\prime }$ does not effect the unitarity. For $\rho _{S}$ a
density matrix on $\mathcal{H}_{S}$ we define the expectation $\langle
.\rangle _{t}$ by 
\begin{equation}
\langle X\rangle _{t}=\mathrm{Tr}\{\rho _{S}\otimes |\Phi _{Q}^{F}\rangle
\langle \Phi _{Q}^{F}|\,U_{t}^{\dagger }(X\otimes 1)U_{t}\}=\mathrm{Tr}%
\{s_{t}X\},  \TCItag{$2.6:8$}
\end{equation}
where $s_{t}$ denotes the effective density matrix on $(S)$ and the second
trace is a partial trace over the system space (the trace over the reservoir
space assumed to be taken already). Now 
\begin{equation}
{\dfrac{d}{dt}}\langle X\rangle _{t}=\mathrm{Tr}\{\rho _{S}\otimes |\Phi
_{Q}^{F}\rangle \langle \Phi _{Q}^{F}|\,{\dfrac{d}{dt}}U_{t}^{\dagger
}(X\otimes 1)U_{t}\},  \TCItag{$2.6:9$}
\end{equation}
so in terms of the effective density matrix $s_{t}$ 
\begin{equation}
{\dfrac{d}{dt}}\mathrm{Tr}\{s_{t}X\}=\mathrm{Tr}\{s_{t}L_{0}(X)\}=\mathrm{Tr}%
\{L_{0}^{\ast }(s_{t})X\},  \TCItag{$2.6:10$}
\end{equation}
where $L_{0}^{\ast }$ denotes the adjoint operation to $L_{0}$ on the dual
of $\mathcal{B}(\mathcal{H}_{S})$. This gives the master equation 
\begin{eqnarray*}
{\dfrac{ds_{t}}{dt}} &=&L_{0}^{\ast }(s_{t})=-\sum_{\omega \in
F}\sum_{j,k=1}^{N(\omega )}\,\{[D_{j}^{\omega }D_{k}^{\omega \dagger
}s_{t}-D_{k}^{\omega \dagger }s_{t}D_{j}^{\omega }]\overline{(g_{j}^{\omega
}|g_{k}^{\omega })}_{Q-}^{\omega -} 
+[D_{j}^{\omega \dagger }D_{k}^{\omega }s_{t}-D_{k}^{\omega
}s_{t}D_{j}^{\omega \dagger }](g_{j}^{\omega }|g_{k}^{\omega })_{Q+}^{\omega
-}
\end{eqnarray*}
\begin{equation}
-[D_{j}^{\omega \dagger }s_{t}D_{k}^{\omega }-s_{t}D_{k}^{\omega
}D_{j}^{\omega \dagger }](g_{j}^{\omega }|g_{k}^{\omega })_{Q-}^{\omega
-}-[D_{j}^{\omega }s_{t}D_{k}^{\omega \dagger }-s_{t}D_{k}^{\omega \dagger
}D_{j}^{\omega }]\overline{(g_{j}^{\omega }|g_{k}^{\omega })}_{Q+}^{\omega
-}\}.  \TCItag{$2.6:11$}
\end{equation}
From the relation (31) we see that $[Y,H_{S}]=0$. If we define the
effective evolution operator by 
\begin{equation}
V_{t}=(e^{{\frac{t}{i\hbar }}H_{S}}\otimes 1)U_{t}  \TCItag{$2.6:12$}
\end{equation}
which satisfies the quantum stochastic differential equation 
\begin{equation}
dV_{t}\equiv ({\frac{1}{i\hbar }}H_{S}e^{{\frac{t}{i\hbar }}H_{S}}\otimes
dt)U_{t}+(e^{{\frac{t}{i\hbar }}H_{S}}\otimes 1)dU_{t}.  \TCItag{$2.6:13$}
\end{equation}
Explicitly this gives 
\begin{eqnarray*}
dV_{t} &\equiv &(e^{{\frac{t}{i\hbar }}H_{S}}\otimes 1)[\sum_{\omega \in
F}\sum_{\phi ,\phi ^{\prime }\in B}^{(\omega _{\phi \phi ^{\prime }}=\omega
)}\{T_{\phi \phi ^{\prime }}\otimes dB_{Q}^{\omega \dagger }(g_{\phi \phi
^{\prime }},t)-T_{\phi \phi ^{\prime }}^{\dagger }\otimes dB_{Q}^{\omega
}(g_{\phi \phi ^{\prime }},t)\} -\{Y+{\dfrac{i}{\hbar }}H_{S}\}dt\ ]\ U_{t}1 \\
&\equiv &[\sum_{\omega \in F}\sum_{\phi ,\phi ^{\prime }\in B}^{(\omega
_{\phi \phi ^{\prime }}=\omega )}\{T_{\phi \phi ^{\prime }}\otimes
dB_{Q}^{\omega \dagger }(g_{\phi \phi ^{\prime }},t)e^{i\omega t}-T_{\phi
\phi ^{\prime }}^{\dagger }\otimes dB_{Q}^{\omega }(g_{\phi \phi ^{\prime
}},t)e^{-i\omega t}\}  -\{{\dfrac{1}{2}}\Gamma +{\dfrac{i}{\hbar }}(H_{S}+{H^{\prime }}_{S})\}dt\
]\ V_{t} . %\TCItag{$2.6:14$}
\end{eqnarray*}
We see that ${H^{\prime }}_{S}$ is a physical addition to the system
Hamiltonian due to the presence of the quantum field. This is precisely the
Lamb shift. Since $Y$ commutes with $H_{S}$ it is enough to compute its
expectations for eigenstates $\phi \in B$; 
\begin{equation}
\langle \phi ,Y\phi \rangle =\sum_{\omega \in F}\sum_{j,k=1}^{N(\omega
)}\{\langle \phi ,D_{j}^{\omega \dagger }D_{k}^{\omega }\phi \rangle \
(g_{j}^{\omega }|g_{k}^{\omega })_{Q+}^{\omega -}\ +\langle \phi
,D_{j}^{\omega }D_{k}^{\omega \dagger }\phi \rangle \ \overline{%
(g_{j}^{\omega }|g_{k}^{\omega })}_{Q-}^{\omega -}\}.  \TCItag{$2.6:15$}
\end{equation}
This is however equivalent to the (complex) shift one calculates using
second order perturbation theory. For example, taking the zero temperature
for simplicity, one calculates in second order shift \cite{19} 
\begin{eqnarray}
Y_{\phi }^{(2)} &=&{\dfrac{1}{i\hbar }}\langle \phi \otimes \Psi _{R},H_{I}{%
\dfrac{1}{H^{(0)}-E_{\phi }-i0^{+}}}H_{I}\,\phi \otimes \Psi _{R}\rangle  
%\TCItag{$2.6:16$} 
\\
&=&\sum_{\omega ,\omega ^{\prime }\in F}\sum_{j}^{N(\omega )}\sum_{j^{\prime
}}^{N(\omega ^{\prime })}\,\int_{-\infty }^{0}d\tau \langle \phi \otimes
\Psi _{R},D_{j}^{\omega }\otimes A(g_{j}^{\omega })  
e^{i(H^{(0)}-E_{\phi })\tau /\hbar }D_{j^{\prime }}^{\omega ^{\prime
}}\otimes A^{\dagger }(g_{j^{\prime }}^{\omega ^{\prime }})\,\phi \otimes
\Psi _{R}\rangle . % \TCItag{$2.6:17$}
\end{eqnarray}
Here we have used the well known identity 
\begin{equation}
\int_{-\infty }^{0}dt\ e^{ixt}={\dfrac{1}{i(x-i0^{+})}}=\pi \delta (x)-i\wp (%
{\dfrac{1}{x}});\ (x\in \mathbb{R}),  \TCItag{$2.6:18$}
\end{equation}
where $\wp $ means that we take the principal part of the integral. Now $%
D_{j}^{\omega }\phi $ is an eigenstate of $H_{S}$ with eigenvalue $E_{\phi
}-\hbar \omega $, so the summation need only be considered over $\omega
=\omega ^{\prime }$ in (2.6:17) above. Therefore we have 
\begin{eqnarray}
Y_{\phi }^{(2)} &=&\sum_{\omega \in F}\sum_{j,k}^{N(\omega )}\,\int_{-\infty
}^{0}d\tau \langle \phi \otimes \Psi _{R},D_{j}^{\omega \dagger }\otimes
A(g_{j}^{\omega })   e^{i(H_{R}-\hbar \omega )\tau /\hbar }\,D_{k}^{\omega }\otimes
A^{\dagger }(g_{k}^{\omega })\,\phi \otimes \Psi _{R}\rangle   \notag \\
&=&\sum_{\omega \in F}\sum_{j,k}^{N(\omega )}\,\langle \phi ,D_{j}^{\omega
\dagger }D_{k}^{\omega }\phi \rangle \,\int_{-\infty }^{0}d\tau \,\langle
\Psi _{R},A(g_{j}^{\omega })A^{\dagger }(S_{\tau }^{\omega }g_{k}^{\omega
})\Psi _{R}\rangle . %\TCItag{$2.6:19$}
\end{eqnarray}
Hence $Y_{\phi }^{(2)}=\langle \phi ,Y\phi \rangle $. The real and imaginary
parts of $Y$ are therefore the linewidth and energy shift as would normally
be calculated using second order perturbation theory, this is true in the
non-vacuum cases also.

\subsection{\textbf{Transition Probabilities}}

Let $P_{t}(\psi |\phi )$ denote the probability that the sytem will be
measured in state $\psi $ at time $t$ if it initially was in state $\phi $,
then 
\begin{equation}
p_{t}(\psi |\phi )=\langle \phi \otimes \Phi _{Q}^{F},U_{t}^{\dagger }(|\psi
\rangle \langle \psi |\otimes 1)U_{t}\,\phi \otimes \Phi _{Q}^{F}\rangle , 
\TCItag{$2.7:1$}
\end{equation}
from theorem 3,3a or 3b we have 
\begin{equation}
{\dfrac{d}{dt}}p_{t}(\psi |\phi )=\langle \phi \otimes \Phi
_{Q}^{F},U_{t}^{\dagger }L_{0}(|\phi \rangle \langle \phi |)\otimes
1U_{t}\,\phi \otimes \Phi _{Q}^{F}\rangle .  \TCItag{$2.7:2$}
\end{equation}
Therefore, setting $t=0$, 
\begin{equation}
{\dfrac{d}{dt}}p_{t}(\psi |\phi )|_{t=0}=\langle \phi ,L_{0}(|\psi \rangle
\langle \psi |)\,\phi \rangle .  \TCItag{$2.7:3$}
\end{equation}
If $\psi =\phi $ we obtain the relation 
\begin{equation}
{\dfrac{d}{dt}}p_{t}(\phi |\phi )|_{t=0}=-\langle \phi ,\Gamma \phi \rangle ,
\TCItag{$2.7:4$}
\end{equation}
while if $\langle \phi ,\psi \rangle =0$ then (112) gives 
\begin{equation*}
{\dfrac{d}{dt}}p_{t}(\psi |\phi )|_{t=0}=\langle \phi ,\Theta (|\psi \rangle
\langle \psi |)\,\phi \rangle ,
\end{equation*}
\begin{eqnarray}
&=&\sum_{\omega \in F}\ \sum_{j,k=1}^{N(\omega )}\{\ \langle \phi
,D_{j}^{\omega \dagger }\psi \rangle \langle \psi ,D_{k}^{\omega }\phi
\rangle (g_{j}^{\omega }|g_{k}^{\omega })_{Q+}^{\omega }\  
+ \langle \phi ,D_{k}^{\omega }\psi \rangle \langle \psi ,D_{j}^{\omega
\dagger }\phi \rangle (g_{j}^{\omega }|g_{k}^{\omega })_{Q-}^{\omega }\ \}, 
%\TCItag{$2.7:5$}
\end{eqnarray}
Using the relation $\int_{-\infty }^{\infty }dt\,e^{ixt}=2\pi \delta (x)$ we
can rewrite this as 
\begin{equation*}
{\dfrac{d}{dt}}p_{t}(\psi |\phi )|_{t=0}=2\pi \sum_{\omega \in F}\
\sum_{j,k=1}^{N(\omega )}\int dk\,\overline{g_{j}^{\omega }}(\mathbf{k}%
)g_{k}^{\omega }(\mathbf{k})\,\delta (\omega (k)-\omega )
\end{equation*}
\begin{equation}
\{\langle \phi ,D_{j}^{\omega \dagger }\psi \rangle \langle \psi
,D_{k}^{\omega }\phi \rangle {\dfrac{q(\mathbf{k})+1}{2}}+\langle \phi
,D_{k}^{\omega }\psi \rangle \langle \psi ,D_{j}^{\omega \dagger }\phi
\rangle {\dfrac{q(\mathbf{k})-1}{2}}\},  \TCItag{$2.7:6$}
\end{equation}
where $q(\mathbf{k})$ is the spectral function associated with $Q$, cf
(125). This is our formulation of the Fermi golden rule for transitions of
the systems state and it corresponds to the usual expressions, cf formulae
(1.21.27a,b) of \cite{2}.

\section{\textbf{THE WEAK COUPLING LIMIT IN QED.}}

As an illustration of our theory we consider the case of quantum
electrodynamics. We stress however that the theory encompasses a wide range
of physical phenomena. For instance exciton models in solid state physics
such as phonon models or the Fr\"{o}lich \cite{15},\cite{16} polaron model
will differ from the following treatment in only minor technical details.
The electromagnetic field acts as reservoir for our system $(S)$ which we
take to consist of a single electron. The electromagnetic field can be
derived from the potential \textbf{A} given by 
\begin{equation}
\mathbf{A}(\mathbf{r})=\sum_{\sigma =1,2}\int \ {\dfrac{d^{3}k}{(2\pi )^{3/2}%
}}\ \ \sqrt{\dfrac{\hbar }{2\epsilon _{o}c|\mathbf{k}|}}\ \{a_{\sigma
}^{\dagger }(\mathbf{k})e^{-i\mathbf{k.r}}+a_{\sigma }(\mathbf{k})e^{i%
\mathbf{k.r}}\}\ \mathbf{\epsilon }^{\sigma }({\hat{\mathbf{k}}}). 
\TCItag{$3:1$}
\end{equation}
Here we consider two transverse polarizations $(\sigma =1,2)$ for each mode $%
\mathbf{k}$. In our notation $\{\mathbf{\epsilon }^{1}({\hat{\mathbf{k}}}),%
\mathbf{\epsilon }^{2}({\hat{\mathbf{k}}}),{\hat{\mathbf{k}}}=|\mathbf{k}%
|^{-1}\mathbf{k}\}$ form a right-handed triad for each $\mathbf{k}$. This
ensures that we are working with the radiation gauge $\nabla .\mathbf{A}=0$.
The operators $a_{\sigma }^{\sharp }(\mathbf{k})$ on the reservoir state
space $\mathcal{H}_{R}$ satisfy Bose commutation relations 
\begin{equation}
\lbrack a_{\sigma }(\mathbf{k}),a_{{\sigma }^{\prime }}^{\dagger }(\mathbf{k}%
^{\prime })]=\delta _{\sigma \sigma ^{\prime }}\delta (\mathbf{k}-\mathbf{%
k^{\prime }}).  \TCItag{$3:2$}
\end{equation}
The total Hamiltonian for the system and reservoir is 
\begin{equation*}
H={\dfrac{1}{2m}}|\mathbf{p}-e\mathbf{A}|^{2}+\Phi (\mathbf{r})+H_{R}
\end{equation*}
\begin{equation}
=H_{S}+H_{R}+H_{I}+{H^{\prime }}_{I},  \TCItag{$3:3$}
\end{equation}
where the unperturbed system Hamiltonian (with potential $\Phi (\mathbf{r})$%
) is 
\begin{equation}
H_{S}={\dfrac{1}{2m}}|\mathbf{p}|^{2}+\Phi (\mathbf{r});  \TCItag{$3:4$}
\end{equation}
\begin{equation}
H_{R}=\sum_{\sigma =1,2}\int d^{3}k\ \ \hbar c|k|\ a_{\sigma }^{\dagger }(%
\mathbf{k})a_{\sigma }(\mathbf{k}),  \TCItag{$3:5$}
\end{equation}
\begin{equation}
H_{I}=-{\dfrac{e}{m}}\sum_{\sigma =1,2}\int d^{3}k\ \{a_{\sigma }^{\dagger }(%
\mathbf{k})e^{-i\mathbf{k.r}}+a_{\sigma }(\mathbf{k})e^{i\mathbf{k.r}}\}\ 
\mathbf{G}^{\sigma }(\mathbf{k}).\mathbf{p},  \TCItag{$3:6$}
\end{equation}
with 
\begin{equation*}
\mathbf{G}^{\sigma }({\hat{\mathbf{k}}})={\dfrac{1}{(2\pi )^{3/2}}}\sqrt{{%
\dfrac{\hbar }{2\epsilon _{o}c|\mathbf{k}|}}}\mathbf{\ \epsilon }^{\sigma }({%
\hat{\mathbf{k}}});
\end{equation*}
and 
\begin{equation*}
{H^{\prime }}_{I}={\dfrac{e^{2}}{2m}}|\mathbf{A}|^{2}.
\end{equation*}

If we rescale the electronic charge as $e\hookrightarrow \lambda e$ we find
that 
\begin{equation*}
H\hookrightarrow H_{S}+H_{R}+\lambda H_{I}+\lambda ^{2}{H^{\prime }}_{I}.
\end{equation*}
In the subsequent analysis we shall drop the term $\lambda ^{2}{H^{\prime }}%
_{I}$ and consider only 
\begin{equation}
H^{(\lambda )}=H_{S}+H_{R}+\lambda H_{I}.  \TCItag{$3:7$}
\end{equation}
It has been established rigorously that this does not effect the final
result in the weak coupling limit. Now the interaction $H_{I}$ given by
(120) has response terms described by the vectors $\theta _{j}^{\sigma }(%
\mathbf{k})={\dfrac{ie}{\hbar m}}e^{-i\mathbf{k.r}}G_{j}^{\sigma }({\hat{%
\mathbf{k}}}).\mathbf{p}$. We assume, as usual, that the unperturbed system
Hamiltonian $H_{S}$ has a complete orthonormal set of of eigenstates $B$. In
the case of the Hydrogen atom, this means that we consider only the bound
states and ignore the effect of the ionized states. In general, $\mathcal{%
H_{S}}$ can be decomposed into complementary subspaces generated by the
discrete, the absolutely continuous and the singular parts of the spectrum
of $H_{S}$. It is enough to prepare the system in the discrete spectrum
subspace to apply our results. It is the standard approach in atomic physics
to study only the behaviour of bound states anyway, so we are justified in
this restriction.

We introduce the test-functions 
\begin{equation}
g_{\phi \phi ^{\prime }}^{\sigma }(\mathbf{k})={\dfrac{ie}{\hbar m}}\langle
\phi |e^{-i\mathbf{k.r}}\mathbf{p}|\phi ^{\prime }\rangle .\mathbf{G}%
^{\sigma }({\hat{\mathbf{k}}}).  \TCItag{$3:8$}
\end{equation}
The interaction $H_{I}$ can be expressed as 
\begin{equation*}
H_{I}=\sum_{\phi ,\phi ^{\prime }\in B}\sum_{\sigma =1,2}\int d^{3}k\
\{a_{\sigma }^{\dagger }(\mathbf{k})g_{\phi \phi ^{\prime }}^{\sigma }(%
\mathbf{k})-a_{\sigma }(\mathbf{k})\overline{g}_{\phi \phi ^{\prime
}}^{\sigma }(\mathbf{k})\}\otimes T_{\phi \phi ^{\prime }}
\end{equation*}
\begin{equation*}
=\sum_{\phi ,\phi ^{\prime }\in B}\sum_{\sigma =1,2}\int d^{3}k\ \{T_{\phi
\phi ^{\prime }}\otimes a_{\sigma }^{\dagger }(\mathbf{k})g_{\phi \phi
^{\prime }}^{\sigma }(\mathbf{k})-T_{\phi ^{\prime }\phi }\otimes a_{\sigma
}(\mathbf{k})\overline{g}_{\phi \phi ^{\prime }}^{\sigma }(\mathbf{k})\}
\end{equation*}
\begin{equation}
=\sum_{\phi ,\phi ^{\prime }\in B}\ \{T_{\phi \phi ^{\prime }}\otimes
A^{\dagger }(g_{\phi \phi ^{\prime }})-T_{\phi \phi ^{\prime }}^{\dagger
}\otimes A(g_{\phi \phi ^{\prime }})\},  \TCItag{$3:9$}
\end{equation}
where $g_{\phi \phi ^{\prime }}=g_{\phi \phi ^{\prime }}^{1}\oplus g_{\phi
\phi ^{\prime }}^{2}\in L^{2}(R^{3})\oplus L^{2}(R^{3})=\mathcal{H}_{R}^{1}$
and $A^{\sharp }$ are the creation/ annihilation operators on $\Gamma _{B}(%
\mathcal{H}_{R}^{1})=\otimes \Gamma _{B}(L^{2}(\mathbb{R}))$; 
\begin{equation*}
A^{\dagger }(f^{1}\oplus f^{2})=\sum_{\sigma =1,2}\int d^{3}kf^{\sigma }(%
\mathbf{k})a_{\sigma }^{\dagger }(\mathbf{k}),
\end{equation*}
\begin{equation}
A(f^{1}\oplus f^{2})=\sum_{\sigma =1,2}\int d^{3}k\overline{f^{\sigma }}(%
\mathbf{k})a_{\sigma }(\mathbf{k}).  \TCItag{$3:10$}
\end{equation}
Our choice of $\mathcal{H}_{R}^{1}$ above for one particle of the reservoir
space is quite natural; namely it consists of wave-functions in the momentum
representation with two transverse polarizations. The state of the reservoir
is in our case determined by the covariance operator $Q$ which we shall now
specify as that of a thermal field at inverse temperature $\beta \rangle 0$,
that is 
\begin{equation*}
Q:\mathcal{H}_{R}^{1}\mapsto \mathcal{H}_{R}^{1}:h_{1}\oplus h_{2}\mapsto {%
\tilde{h}}_{1}\oplus {\tilde{h}}_{2}
\end{equation*}
\begin{equation}
\mathrm{with\ }{\tilde{h}}_{\sigma }(\mathbf{k})=q(c|\mathbf{k}|)h_{\sigma }(%
\mathbf{k}),  \TCItag{$3:11$}
\end{equation}
where $q(\omega )=\coth {\dfrac{\beta \hbar \omega }{2}}.$ With $\omega
_{\phi \phi ^{\prime }}=(E_{\phi ^{\prime }}-E_{\phi })/\hbar \in F$,we
define 
\begin{equation*}
S_{t}^{\omega _{\phi \phi ^{\prime }}}:\mathcal{H}_{R}^{1}\mapsto \mathcal{H}%
_{R}^{1}:h_{1}\oplus h_{2}\mapsto {\tilde{h}}_{1}\oplus {\tilde{h}}_{2}
\end{equation*}
with 
\begin{equation}
{\tilde{h}}_{\sigma }(\mathbf{k})=e^{i(c|\mathbf{k}|-\omega _{\phi \phi
^{\prime }})t}h_{\sigma }(\mathbf{k}).  \TCItag{$3:12$}
\end{equation}
In this setup we have allowed for the most general coupling, that is where
all the fundamental frequencies $F=\{\omega _{\phi \phi ^{\prime }}:\phi
,\phi ^{\prime }\in B\}$ are to be considered. This set is always degenerate
in general; however it is important to consider two classes of degeneracy
arising. The first is the secular class; these are the situations in which
degeneracies always arise regardless of the spectrum $\{E_{\phi }:\phi \in
B\}$ of $H_{S}$; they are the pairs $(\phi ,\phi ^{\prime })$ and $(\psi
,\psi ^{\prime })$ which have $\omega _{\phi \phi ^{\prime }}=\omega _{\psi
\psi ^{\prime }}$ due to one of the following reasons 
\begin{equation*}
1.\ \ \phi =\phi ^{\prime }=\psi =\psi ^{\prime },
\end{equation*}
\begin{equation*}
2.\ \ \ \phi =\phi ^{\prime },\psi =\psi ^{\prime };\ (\phi \neq \psi ),
\end{equation*}
\begin{equation}
3.\ \ \ \phi =\psi ,\phi ^{\prime }=\psi ^{\prime };\ (\phi \neq \phi
^{\prime }).  \TCItag{$3:13$}
\end{equation}
Any solution to the equation $\omega _{\phi \phi ^{\prime }}=\omega _{\psi
\psi ^{\prime }}$, or equivalently $E_{\phi }-E_{\phi ^{\prime }}=E_{\psi
}-E_{\psi ^{\prime }}$, not of the secular type shall be called an
extraneous solution. The extraneous solutions are of course dependent on the
spectrum of $H_{S}$. It is standard procedure in the physical literature to
assume that such possibilities do not arise however this is a requirement on 
$H_{S}$ which cannot be made in many important examples. For a particle in a
rectangular box, apart from the natural degeneracies arising if the ratios
of the sides are rational, we also have to consider the fact that the
contribution to the energy for the mode of vibration $n_{i}$ along the $i^{%
\mathrm{th}}$-axis is proportional to $n_{i}^{2}$, this means solving the
Diophantine equations for the harmonics 
\begin{equation*}
n_{i}^{2}-m_{i}^{2}={n^{\prime }}_{i}^{2}-{\ m^{\prime }}_{i}^{2}.
\end{equation*}
For the Hydrogen atom we have, apart from the spherical harmonical
degeneracies, to consider the integer solutions to the Diophantine equations 
\begin{equation*}
{\dfrac{1}{n^{2}}}-{\dfrac{1}{m^{2}}}={\dfrac{1}{{n^{\prime }}^{2}}}-{\dfrac{%
1}{{m^{\prime }}^{2}}}
\end{equation*}
for the principal atomic numbers. After simple manipulations this leads to
the study of the intersection of the algebraic projective curve in $\mathbb{R%
}^{4}$; 
\begin{equation*}
x_{1}^{2}x_{3}^{2}x_{4}^{2}-x_{2}^{2}x_{3}^{2}x_{4}^{2}-x_{1}^{2}x_{2}^{2}x_{4}^{2}+x_{1}^{2}x_{2}^{2}x_{3}^{2}=0
\end{equation*}
with the lattice of positive integers. In the weak coupling limit we obtain
the quantum stochastic differential equation 
\begin{equation*}
dU_{t}=[\sum_{\omega \in F}\sum_{\phi ,\phi ^{\prime }\in B}^{(\omega _{\phi
\phi ^{\prime }}=\omega )}\{T_{\phi \phi ^{\prime }}\otimes dB_{Q}^{\omega
\dagger }(g_{\phi \phi ^{\prime }},t)+T_{\phi \phi ^{\prime }}^{\dagger
}\otimes dB_{Q}^{\omega }(g_{\phi \phi ^{\prime }},t)\}+Ydt\ ]\ U_{t}
\end{equation*}
where
\begin{equation}
Y=\sum_{\omega \in F}\sum_{\phi ,\phi ^{\prime },\psi ,\psi ^{\prime }\in
B}^{(\omega _{\phi \phi ^{\prime }}=\omega =\omega _{\psi \psi ^{\prime
}})}\ \ [T_{\phi \phi ^{\prime }}^{\dagger }T_{\psi \psi ^{\prime }}(g_{\phi
\phi ^{\prime }}|g_{\psi \psi ^{\prime }})_{Q+}^{\omega -}+T_{\phi \phi
^{\prime }}T_{\psi \psi ^{\prime }}^{\dagger }\overline{(g_{\phi \phi
^{\prime }}|g_{\psi \psi ^{\prime }})_{Q-}^{\omega -}}];  \TCItag{$3:14$}
\end{equation}
with $U_{0}=1$. However using the fact that $T_{\phi \phi ^{\prime
}}^{\dagger }T_{\psi \psi ^{\prime }}=\langle \phi ,\psi \rangle T_{\phi
^{\prime }\psi ^{\prime }}$ etc., we may write $Y$ as 
\begin{equation}
Y=\sum_{\omega \in F}\sum_{\phi ,\psi ,\phi ^{\prime }\in B}^{(\omega
=\omega _{\phi \phi ^{\prime }}=\omega _{\psi \phi ^{\prime }})}\ \{(g_{\phi
^{\prime }\phi }|g_{\phi ^{\prime }\psi })_{Q+}^{(-\omega )-}+\overline{%
(g_{\phi \phi ^{\prime }}|g_{\psi \phi ^{\prime }})_{Q-}^{\omega -}}\}\
T_{\phi \psi }.  \TCItag{$3:15$}
\end{equation}
In the summation we consider only $\phi $ and $\psi $ for which there exists
a $\phi ^{\prime }$ so that $\omega _{\phi \phi ^{\prime }}=\omega _{\psi
\phi ^{\prime }}$, however this is equivalent to demanding that $\omega
_{\phi \psi }=0$ as we always have the identity $\omega _{\phi \psi }=\omega
_{\phi \phi ^{\prime }}-\omega _{\psi \phi ^{\prime }}$. Therefore $Y$ is a
linear combination of terms $T_{\phi \psi }$ with $\omega _{\phi \psi }=0$
and this, in particular, implies that $Y$ commutes with $H_{S}$. It is
natural in light of this to write $Y$ as 
\begin{equation}
Y=\sum_{\phi ,\psi \in B}^{(\omega _{\phi \psi }=0)}\ y_{\phi \psi }\
T_{\phi \psi }=\sum_{\phi ,\psi \in B}^{(E_{\phi }=E_{\psi })}\ \ y_{\phi
\psi }\ T_{\phi \psi };  \TCItag{$3:16$}
\end{equation}
where 
\begin{equation}
y_{\phi \psi }=\ \sum_{\phi ^{\prime }\in B}\ \{(g_{\phi ^{\prime }\phi
}|g_{\phi ^{\prime }\psi })_{Q+}^{\omega _{\phi ^{\prime }\phi }-}+\overline{%
(g_{\phi \phi ^{\prime }}|g_{\psi \phi ^{\prime }})_{Q-}^{\omega _{\phi \phi
^{\prime }}-}}\}.  \TCItag{$3:17$}
\end{equation}
Now the master equation associated with this problem is from (105) 
\begin{equation}
{\dfrac{ds_{t}}{dt}}=L_{0}^{\ast }(s_{t})=-(Ys_{t}+s_{t}Y^{\dagger })
+\sum_{\omega \in F}\sum_{\phi ,\phi ^{\prime },\psi ,\psi ^{\prime }\in
B}^{(\omega =\omega _{\phi \phi ^{\prime }}=\omega _{\psi \psi ^{\prime
}})}[T_{\phi \phi ^{\prime }}^{\dagger }s_{t}T_{\psi \psi ^{\prime
}}(g_{\phi \phi ^{\prime }}|g_{\psi \psi ^{\prime }})_{Q+}^{\omega }+T_{\phi
\phi ^{\prime }}s_{t}T_{\psi \psi ^{\prime }}^{\dagger }(g_{\psi \psi
^{\prime }}|g_{\phi \phi ^{\prime }})_{Q-}^{\omega }\ ].  \TCItag{$3:18$}
\end{equation}
In order to find the general expression for $H_{S}^{\prime }$, we see that 
\begin{equation*}
y_{\phi \psi }=\sum_{\phi ^{\prime }\in B}\int_{-\infty }^{0}d\tau \{\langle
g_{\phi ^{\prime }\phi },S_{\tau }^{\omega _{\phi ^{\prime }\phi }}{\dfrac{%
Q+1}{2}}g_{\phi ^{\prime }\psi }\rangle +\overline{\langle g_{\phi \phi
^{\prime }},S_{\tau }^{\omega _{\phi \phi ^{\prime }}}{\dfrac{Q-1}{2}}%
g_{\psi \phi ^{\prime }}\rangle }\}
\end{equation*}
\begin{eqnarray*}
&=&\sum_{\phi ^{\prime }\in B}\int_{-\infty }^{0}d\tau \sum_{\sigma
=1,2}\int d^{3}k\,\{\overline{g_{\phi ^{\prime }\phi }^{\sigma }}(\mathbf{k}%
)g_{\phi ^{\prime }\psi }^{\sigma }(\mathbf{k})e^{-ic|(\mathbf{k}|t}{\dfrac{%
q(c|\mathbf{k}|)+1}{2}}
+g_{\phi \phi ^{\prime }}^{\sigma }(\mathbf{k})\overline{g_{\psi \phi
^{\prime }}^{\sigma }}(\mathbf{k})e^{ic|\mathbf{k}|t}{\dfrac{q(c|\mathbf{k}%
|)-1}{2}}\}e^{i\omega _{\phi ^{\prime }\phi }t}.
\end{eqnarray*}
But using $\overline{g_{\phi \phi ^{\prime }}^{\sigma }}(\mathbf{k}%
)=-g_{\phi ^{\prime }\phi }^{\sigma }(-\mathbf{k})$ we have 
\begin{eqnarray}
y_{\phi \psi } &=&\sum_{\phi ^{\prime }\in B}\int_{-\infty }^{0}d\tau
\sum_{\sigma =1,2}\int d^{3}k\,g_{\phi \phi ^{\prime }}^{\sigma }(\mathbf{k})%
\overline{g_{\psi \phi ^{\prime }}^{\sigma }}(\mathbf{k})  
\{e^{-ic|\mathbf{k}|t}{\dfrac{q(c|\mathbf{k}|)+1}{2}}+e^{ic|\mathbf{%
k}|t}{\dfrac{q(c|\mathbf{k}|)-1}{2}}\}e^{i\omega _{\phi ^{\prime }\phi }t} 
\notag \\
&=&{\dfrac{e^{2}}{\hbar ^{2}m^{2}}}\sum_{\phi ^{\prime }\in B}\int_{-\infty
}^{0}d\tau \sum_{\sigma =1,2}\int d^{3}k\sum_{j,j^{\prime }=1,2,3}\langle
\phi |e^{-i\mathbf{k.r}}p_{j}|\phi ^{\prime }\rangle \langle \phi ^{\prime
}|e^{i\mathbf{k.r}}p_{j^{\prime }}|\psi \rangle    \notag \\
&&\times G_{j}^{\sigma }(\mathbf{k})G_{j^{\prime }}^{\sigma }(\mathbf{k}%
)\{e^{-ic|\mathbf{k}|t}{\dfrac{q(c|\mathbf{k}|)+1}{2}}+e^{ic|\mathbf{k}|t}{%
\dfrac{q(c|\mathbf{k}|)-1}{2}}\}e^{i\omega _{\phi ^{\prime }\phi }t}  \notag
\\
&=&{\dfrac{e^{2}}{\hbar ^{2}m^{2}}}\sum_{\phi ^{\prime }\in B}\int_{-\infty
}^{0}d\tau \sum_{\sigma =1,2}\int d^{3}k\sum_{j,j^{\prime }=1,2,3}\langle
\phi |e^{{\dfrac{t}{i\hbar }}E_{\phi }}e^{-i\mathbf{k.r}}p_{j}e^{{\dfrac{-t}{%
i\hbar }}H_{S}}e^{i\mathbf{k.r}}p_{j^{\prime }}|\psi \rangle   \notag \\
&&\times G_{j}^{\sigma }(\mathbf{k})G_{j^{\prime }}^{\sigma }(\mathbf{k}%
)\{e^{-ic|\mathbf{k}|t}{\dfrac{q(c|\mathbf{k}|)+1}{2}}+e^{ic|\mathbf{k}|t}{%
\dfrac{q(c|\mathbf{k}|)-1}{2}}\}.  %\TCItag{$3:19$}
\end{eqnarray}
We remark that the effect of the response term is as follows; from the
commutation relations of $\mathbf{r}$ and $\mathbf{p}$ we have that 
\begin{equation}
e^{-i\mathbf{k.r}}e^{{\dfrac{t}{i\hbar }}H_{S}}e^{i\mathbf{k.r}}=e^{-i%
\mathbf{k.r}}\exp {\dfrac{t}{i\hbar }}({\dfrac{|\mathbf{p}|^{2}}{2m}}%
+V(r))e^{i\mathbf{k.r}}=\exp {\dfrac{t}{i\hbar }}({\dfrac{|\mathbf{p}+\hbar 
\mathbf{k}|^{2}}{2m}}+V(r)),  \TCItag{$3:20$}
\end{equation}
that is $H_{S}$ is replacd by $H_{S}+{\dfrac{\hbar \mathbf{k}.\mathbf{p}}{m}}%
+{\dfrac{\hbar ^{2}|\mathbf{k}|^{2}}{2m}}.$ Now the $\mathbf{k}$ dependence
in the above expression prevents us from using the well-known isotropic
identity 
\begin{equation}
\sum_{\sigma =1,2}\int_{|\mathbf{k}|=\omega /c}d^{2}\hat{k}\,G_{j}^{\sigma }(%
\mathbf{k})G_{j^{\prime }}^{\sigma }(\mathbf{k})={\dfrac{1}{(2\pi )^{3}}}{%
\dfrac{\hbar }{2\epsilon _{0}\omega }}{\dfrac{8\pi }{3}}\delta _{j,j^{\prime
}},  \TCItag{$3:21$}
\end{equation}
to calculate $y_{\phi \psi }$ as in the dipole approximation. Note that the
Lamb shift and the damping coefficients are effected by inclusion of the
response terms.

The complex shift $Y_{\phi \phi }={\dfrac{1}{2}}\Gamma _{\phi }+{\dfrac{i}{%
\hbar }}E_{\phi }^{\prime }$, giving the linewidth $\Gamma _{\phi }$ and
energy shift $E_{\phi }^{\prime }$ for a state $\phi \in B$ can the be
written as 
\begin{eqnarray}
Y_{\phi \phi } &=&{\dfrac{e^{2}}{2i\hbar m^{2}}}\sum_{\sigma =1,2}\int
d^{3}k\sum_{j,j^{\prime }=1,2,3}  
\langle \phi |p_{j}[{\dfrac{q(c|\mathbf{k}|)+1}{\mathcal{D}^{+}(\mathbf{k}%
)-i0^{+}}}+{\dfrac{q(c|\mathbf{k}|)-1}{\mathcal{D}^{-}(\mathbf{k})-i0^{+}}}%
]p_{j^{\prime }}|\phi \rangle G_{j}^{\sigma }(\mathbf{k})G_{j^{\prime
}}^{\sigma }(\mathbf{k}),
\end{eqnarray}
where the denominators in the above expression are 
\begin{equation}
\mathcal{D}^{\pm }(\mathbf{k})=H_{S}+{\dfrac{\hbar \mathbf{k}.\mathbf{p}}{m}}%
+{\dfrac{\hbar ^{2}|\mathbf{k}|^{2}}{2m}}\pm \hbar c|\mathbf{k}|-E_{\phi }. 
\TCItag{$3:23$}
\end{equation}
This expression has been derived in \cite{20} for the zero temperature case;
see however \cite{21}.

{\large APPENDIX. A:}\textbf{\ The Traditional Derivation of the Master
Equation.}

For the sake of comparison we give the standard arguments used in the
derivation of the master equation. This section follows closely the
development of Louisell \cite{2}. The interaction is taken to be of the form 
\begin{equation*}
H_{I}=\sum_{j}D_{j}\otimes F_{j},
\end{equation*}
where $D_{j}$ and $F_{j}$ act nontrivially on the system and reservoir
spaces respectively. We assume that $D_{j}$ evolves harmonically in time
under the free evolution with frequency $\omega _{j}$ We assume that at time 
$t$ the system and reservoir are uncoupled, that is the density operator $%
\rho (t)$ at time zero factors as 
\begin{equation*}
\rho (0)=\rho _{0}^{(S)}\otimes \rho ^{(R)}.
\end{equation*}
No subscript is required for $\rho ^{(R)}$ as we assume that it is invariant
under the free-evolution (in particular this is true for the choice of a
thermal state $\rho ^{(R)}=e^{-\beta (H_{R}^{1}-\mu )}/\mathrm{Tr}e^{-\beta
(H_{R}^{1}-\mu )}$). We define the reduced system state at time $t$ in the
interaction dynamics to be the density operator 
\begin{equation*}
s_{t}=\mathrm{Tr}_{\mathcal{H}_{R}}\{U_{t}^{(\lambda )}(\rho
_{0}^{(S)}\otimes \rho ^{(R)})U_{t}^{(\lambda )\dagger }\}.
\end{equation*}
The iterated series expansion of $s_{t}$, truncated to second order, is; 
\begin{equation*}
s_{t} =s_{0}+{\dfrac{1}{i\hbar }}\int_{0}^{t}dt_{1}\mathrm{Tr}_{\mathcal{H}%
_{R}}[v_{t_{1}}^{(0)}(H_{I}),s_{0}\otimes \rho ^{(R)}] +{\dfrac{1}{(i\hbar )^{2}}}\int_{0}^{t}dt_{1}\int_{0}^{t_{1}}dt_{2}\mathrm{%
Tr}_{\mathcal{H}_{R}}[v_{t_{1}}^{(0)}(H_{I}),[v_{t_{2}}^{(0)}(H_{I}),s_{0}%
\otimes \rho ^{(R)}]],
\end{equation*}
where we have set $\lambda =1$. Substituting in for the potential $H_{I}$ we
find 
\begin{eqnarray*}
s_{t} &=&s_{0}+\sum_{j}\int_{0}^{t}dt_{1}\langle
v_{t_{1}}^{(0)}(F_{j})\rangle _{R}e^{-i\omega _{j}t_{1}}[D_{j},s_{0}] \\
&&+\sum_{j,k}\int_{0}^{t}dt_{1}\int_{0}^{t_{1}}dt_{2}e^{-i(\omega
_{j}t_{1}+\omega _{k}t_{2})}\{[D_{j}D_{k}s_{0}-D_{k}s_{0}D_{j}]\langle
v_{t_{1}}^{(0)}(F_{j})v_{t_{2}}^{(0)}(F_{k})\rangle _{R} -[D_{j}s_{0}D_{k}-s_{0}D_{k}D_{j}]\langle
v_{t_{2}}^{(0)}(F_{k})v_{t_{1}}^{(0)}(F_{j})\rangle _{R}\},
\end{eqnarray*}
where $\langle .\rangle _{R}=\mathrm{Tr}_{\mathcal{H}_{R}}[\rho ^{(R)}.]$.
Due to the invariance of the reservoir fields under the free evolution we
have that 
\begin{equation*}
\langle v_{t}^{(0)}(F_{j})\rangle _{R}=\langle F_{j}\rangle _{R};
\end{equation*}
\begin{equation*}
\langle v_{t}^{(0)}(F_{j})v_{s}^{(0)}(F_{k})\rangle _{R}=\langle
v_{t-s}^{(0)}(F_{j})F_{k}\rangle _{R}.
\end{equation*}
Therefore, if we let $\tau =t_{1}-t_{2},y=t_{2}$ then we obtain 
\begin{eqnarray*}
s_{t} &=&s_{0}+\sum_{j}\langle F_{j}\rangle
_{R}[D_{j},s_{0}]\int_{0}^{t}e^{-i\omega _{j}y}dy \\
&&+\sum_{j,k}\int_{0}^{t}dy\,e^{-i(\omega _{j}+\omega
_{k})y}\,\int_{0}^{t-y}d\tau \,e^{-i\omega _{j}\tau
}\{[D_{j}D_{k}s_{0}-D_{k}s_{0}D_{j}]\langle v_{\tau
}^{(0)}(F_{j})F_{k}\rangle _{R} -[D_{j}s_{0}D_{k}-s_{0}D_{k}D_{j}]\langle F_{k}v_{\tau
}^{(0)}(F_{j})\rangle _{R}\}.
\end{eqnarray*}
The approximation procedure is based on the following four steps; \textit{%
step I}. One postulates that the contributions coming from the sum of all
terms higher that second order in the iterated series are negligible. 
\textit{step II}. One postulates a finite autocorrelation time $\tau _{c}$
such that 
\begin{equation*}
\langle v_{\tau }^{(0)}(F_{j})F_{k}\rangle _{R}=0=\langle F_{j}v_{\tau
}^{(0)}(F_{k})\rangle _{R},
\end{equation*}
whenever $|\tau |\rangle \tau _{c}.$ Thus for $t\rangle \rangle \tau _{c}$
one may replace the upper limit of the $\tau $-integral by $+\infty $. This
gives 
\begin{equation*}
s_{t}=s_{0}+\sum_{j}\langle F_{j}\rangle _{R}[D_{j},s_{0}]I^{t}(\omega _{j})
\end{equation*}
\begin{equation*}
+\sum_{j,k}\,%
\{[D_{j}D_{k}s_{0}-D_{k}s_{0}D_{j}]w_{j,k}^{+}-[D_{j}s_{0}D_{k}-s_{0}D_{k}D_{j}]w_{k,j}^{-}\}I^{t}(\omega _{j}+\omega _{k}),
\end{equation*}
where 
\begin{eqnarray*}
w_{j,k}^{+} &=&\int_{0}^{\infty }e^{-i\omega _{j}\tau }\langle v_{\tau
}^{(0)}(F_{j})F_{k}\rangle _{R}d\tau  \\
w_{k,j}^{-} &=&\int_{0}^{\infty }e^{-i\omega _{j}\tau }\langle F_{k}v_{\tau
}^{(0)}(F_{j})\rangle _{R}d\tau 
\end{eqnarray*}
and 
\begin{equation*}
I^{t}(\omega )=\int_{0}^{t}e^{-i\omega y}dy.
\end{equation*}
\textit{Step III}. For $t$ large with respect to $\tau _{c}$ one makes the
replacement 
\begin{equation*}
I^{t}(\omega )\hookrightarrow t\delta (\omega ).
\end{equation*}

\textit{Step IV}. One postulates that the formulas deduced under the
previous assumptions when $t$ is large with respect to $\tau _{c}$ hold also
in the limit $t\rightarrow 0$; this gives 
\begin{equation*}
{\dfrac{ds}{dt}}|_{0}=\lim_{t\rightarrow 0}{\dfrac{s_{t}-s_{0}}{t}}
\end{equation*}
\begin{equation*}
=\sum_{j}^{\omega _{j}=0}[D_{j},s_{0}]\langle F_{j}\rangle
_{R}+\sum_{j,k}^{\omega _{j}+\omega
_{k}=0}\,%
\{[D_{j}D_{k}s_{0}-D_{k}s_{0}D_{j}]w_{j,k}^{+}-[D_{j}s_{0}D_{k}-s_{0}D_{k}D_{j}]w_{k,j}^{-}\}.
\end{equation*}
The assumptions leading to this equation have a decidedly ad hoc nature,
especially those introduced in steps III and IV. The replacement for $I^{t}$%
, put in by hand, in step III is precisely what is needed to allow the limit
to be taken easily.

It is instructive to calculate explicitly the master equation in a
particular case. We consider as reservoir a free Bose gas at inverse
temperature $\beta $ and fugacity $z=e^{\beta \mu }$. This can be described
by the quasi-free state $\varphi _{Q}$, on $L^{2}(\mathbb{R}^{n})$ for
example, characterized by 
\begin{equation*}
\langle A^{\dagger }(f)A(g)\rangle _{R}\equiv \varphi _{Q}(A^{\dagger
}(f)A(g))=\langle f,{\dfrac{Q-1}{2}}g\rangle ,
\end{equation*}
where 
\begin{equation*}
Q={\dfrac{1+ze^{-\beta H_{R}^{1}}}{1-ze^{-\beta H_{R}^{1}}}}=\coth {\dfrac{%
\beta }{2}}(H_{R}^{1}-\mu ).
\end{equation*}
We may take $H_{R}^{1}$ to be for instance $-\Delta $.

We may write the interaction $H_{I}$ of (31) in the form we are
considering with the notations 
\begin{equation*}
H_{I}=i\hbar \sum_{j}\{D_{j}\otimes A^{\dagger }(g_{j})-h.c.\}\equiv i\hbar
\sum_{(j,\alpha )}D_{(j,\alpha )}\otimes F_{(j,\alpha )},
\end{equation*}
where we have a summation also over an index $\alpha \in \{0,1\}$ with the
notations 
\begin{equation*}
D_{(j,0)}=D_{j},\,D_{(j,1)}=-D_{j}^{\dagger };\,\,F_{(j,0)}=A^{\dagger
}(g_{j}),\,F_{(j,1)}=A(g_{j});
\end{equation*}
and consequently 
\begin{equation*}
\omega _{(j,0)}=\omega _{j};\,\omega _{(j,1)}=-\omega _{j}.
\end{equation*}
We then have 
\begin{eqnarray*}
w_{(j0),(k1)}^{+} &=&\int_{0}^{\infty }d\tau \,e^{-i\omega _{j}\tau }\langle
v_{\tau }^{(0)}(A^{\dagger }(g_{j}))A(g_{k})\rangle _{R}=\int_{0}^{\infty
}d\tau \,\varphi _{Q}(A^{\dagger }(S_{\tau }^{\omega _{j}}g_{j})A(g_{k})) \\
&=&\int_{0}^{\infty }d\tau \,\langle S_{\tau }^{\omega _{j}}g_{j},{\dfrac{Q-1%
}{2}}g_{k}\rangle =\overline{(g_{j}|g_{k})}_{Q-}^{\omega _{j}-},
\end{eqnarray*}
Similarly, using the CCR, we find 
\begin{equation*}
w_{(j1),(k0)}^{+}=\int_{0}^{\infty }d\tau \,e^{i\omega _{j}\tau }\langle
v_{\tau }^{(0)}(A(g_{j}))A^{\dagger }(g_{k})\rangle
_{R}=(g_{j}|g_{k})_{Q+}^{\omega _{j}-},
\end{equation*}
\begin{eqnarray*}
w_{(k0),(j1)}^{-} &=&(g_{j}|g_{k})_{Q-}^{\omega _{j}-}, \\
w_{(j1),(k0)}^{-} &=&\overline{(g_{j}|g_{k})}_{Q+}^{\omega _{j}-},
\end{eqnarray*}
while $w_{(j\epsilon ),(j^{\prime }\epsilon ^{\prime })}^{\pm }=0$ if $%
\epsilon =\epsilon ^{\prime }$ as we have $\langle A(f)A(g)\rangle
_{R}=0=\langle A^{\dagger }(f)A^{\dagger }(g)\rangle _{R}.$ We note that $%
\langle F_{(j,\alpha )}\rangle _{R}=0$ in all cases.

The master equation then reads 
\begin{eqnarray*}
{\dfrac{ds}{dt}}|_{0} &=&\sum_{j,k;\alpha ,\alpha ^{\prime }}^{(-1)^{\alpha
}\omega _{j}+(-1)^{\alpha ^{\prime }}\omega _{k}=0}\,\{[D_{j,\alpha
}D_{k,\alpha ^{\prime }}s_{0}-D_{k,\alpha ^{\prime }}s_{0}D_{j\alpha
}]w_{(j\alpha ),(k\alpha ^{\prime })}^{+} \\
&&-[D_{j,\alpha }s_{0}D_{k,\alpha ^{\prime }}-s_{0}D_{k,\alpha ^{\prime
}}D_{j,\alpha }]w_{(k\alpha ^{\prime }),(j\alpha )}^{-}\} \\
&=&-\sum_{j,k}^{\omega _{j}-\omega _{k}=0}\{[D_{j}D_{k}^{\dagger
}s_{0}-D_{k}^{\dagger }s_{0}D_{j}]w_{(j0),(k1)}^{+}+[D_{j}^{\dagger
}D_{k}s_{0}-D_{k}s_{0}D_{j}^{\dagger }]w_{(j1),(k0)}^{+} \\
&&-[D_{j}^{\dagger }s_{0}D_{k}-s_{0}D_{k}D_{j}^{\dagger
}]w_{(k0),(j1)}^{-}-[D_{j}s_{0}D_{k}^{\dagger }-s_{0}D_{k}^{\dagger
}D_{j}]w_{(k1),(j0)}^{-}\},
\end{eqnarray*}
or writing in our notations (and employing the relabeling in terms of the
frequency degeneracies as in (31)) 
\begin{eqnarray*}
{\dfrac{ds}{dt}}|_{0} &=&-\sum_{\omega \in F}\sum_{j,k=1}^{N(\omega
)}\,\{[D_{j}^{\omega }D_{k}^{\omega \dagger }s_{0}-D_{k}^{\omega \dagger
}s_{0}D_{j}^{\omega }]\overline{(g_{j}^{\omega }|g_{k}^{\omega })}%
_{Q-}^{\omega -} \\
&&+[D_{j}^{\omega \dagger }D_{k}^{\omega }s_{0}-D_{k}^{\omega
}s_{0}D_{j}^{\omega \dagger }](g_{j}^{\omega }|g_{k}^{\omega })_{Q+}^{\omega
-} \\
&&-[D_{j}^{\omega \dagger }s_{0}D_{k}^{\omega }-s_{0}D_{k}^{\omega
}D_{j}^{\omega \dagger }](g_{j}^{\omega }|g_{k}^{\omega })_{Q-}^{\omega
-}-[D_{j}^{\omega }s_{0}D_{k}^{\omega \dagger }-s_{0}D_{k}^{\omega \dagger
}D_{j}^{\omega }]\overline{(g_{j}^{\omega }|g_{k}^{\omega })}_{Q+}^{\omega
-}\}.
\end{eqnarray*}
But this is exactly ${\dfrac{ds_{t}}{dt}}|_{0}=L_{0}^{\ast }(s_{0})$, where $%
L_{0}^{\ast }$ is given as before.

{\large APPENDIX\ B:}\textbf{\ The Convergence of the Collective Processes
to the Noise}

\textbf{Processes}

The mathematical theory behind the weak coupling limit developed in [1] and
subsequent papers is the following. We estimate the behaviour as $\lambda
\rightarrow 0$ of matrix elements of $U_{t/\lambda ^{2}}^{(\lambda )}$ with
respect to collective coherent vectors, that is vectors of the form $%
B_{t^{(1)}/\lambda ^{2}}^{(\omega ,\lambda )}(f^{(1)})...B_{t^{(n)}/\lambda
^{2}}^{(\omega ,\lambda )}(f^{(n)})\Psi _{R}$. This involves substituting $%
v_{t}(H_{I})$ into the series
expansion for $U_{t/\lambda ^{2}}^{(\lambda )}$ and examining each
of the terms arising. The deep analysis of [1] shows that each term upon
normal ordering leads to two classes of terms: relevant ones (type I) and
negligible ones (type II). The type I terms are exactly those put into
normal order by commuting \textit{time consecutive} pairs of reservoir
variables, the type II terms account for all others. Following this
resummation it is shown that the type II terms give vanishing contribution
in the limit $\lambda \rightarrow 0$ while the explicit limit for the type I
terms is calculated; uniform convergence is established, the main technical
device used here is one of various generalizations of the Pul\'{e}
inequality [3].

The independence of the noise processes for different frequencies follows
from the next two lemmas.

\bigskip

\textbf{Lemma 1:} \textit{for each }$\omega \in F$\textit{\ let }$f_{\omega
},f_{\omega }^{\prime }\in K_{\omega }$\textit{\ and }$S_{\omega },T_{\omega
},S_{\omega }^{\prime },T_{\omega }^{\prime }\in R$\textit{\ then } 
\begin{eqnarray*}
\lim_{\lambda \rightarrow 0} &\langle &\lambda \int_{S_{\omega }/\lambda
^{2}}^{T_{\omega }/\lambda ^{2}}S_{u}^{\omega }f_{\omega }du,\lambda \int_{{%
S^{\prime }}_{\omega }/\lambda ^{2}}^{{T^{\prime }}_{\omega }/\lambda
^{2}}S_{v}^{\omega }{f^{\prime }}_{\omega }dv\rangle  \\
&=&\sum_{\omega \in F}\langle \chi _{\lbrack S_{\omega },T_{\omega }]},\chi
_{\lbrack {S^{\prime }}_{\omega },{T^{\prime }}_{\omega }]}\rangle _{L^{2}(%
\mathbb{R})}(f_{\omega }|{f^{\prime }}_{\omega })_{\omega } \\
&=&(\oplus _{\omega \in F}(\chi _{\lbrack S_{\omega },T_{\omega }]}\otimes
f_{\omega })|\oplus _{\omega ^{\prime }\in F}(\chi _{\lbrack {S^{\prime }}%
_{\omega ^{\prime }},{T^{\prime }}_{\omega ^{\prime }}]}\otimes {f^{\prime }}%
_{\omega ^{\prime }}).
\end{eqnarray*}

\begin{proof}
the left hand side of (B.1) can be written as a sum over $\omega ,\omega
^{\prime }\in F$ of terms 
\begin{equation*}
\lim_{\lambda \rightarrow 0}\int_{S_{\omega }}^{T_{\omega }}du\int_{({%
S^{\prime }}_{\omega ^{\prime }}-u)/\lambda ^{2}}^{({T^{\prime }}_{\omega
^{\prime }}-u)/\lambda ^{2}}du^{\prime }\langle f_{\omega },S_{u^{\prime
}}^{\omega }{f^{\prime }}_{\omega ^{\prime }}\rangle e^{i(\omega -\omega
^{\prime })u/\lambda ^{2}};
\end{equation*}
by the Riemann-Lesbegue Lemma the terms $\omega \neq \omega ^{\prime }$
vanish while the $\omega =\omega ^{\prime }$ terms converge by inspection to 
\begin{equation*}
\langle \chi _{\lbrack S_{\omega },T_{\omega }]},\chi _{\lbrack {S^{\prime }}%
_{\omega },{T^{\prime }}_{\omega }]}\rangle _{L^{2}(\mathbb{R})}(f_{\omega }|%
{f^{\prime }}_{\omega })_{\omega }
\end{equation*}
\end{proof}

\bigskip

\textbf{Lemma 2:}\textit{\ for }$n\in N$\textit{\ let }$f_{\omega }^{(k)}\in
K_{\omega },x_{\omega }^{(k)}\in R,S_{\omega }^{(k)}\langle T_{\omega
}^{(k)},$\textit{\ for }$1\leq k\leq n$\textit{\ and each }$\omega \in F$%
\textit{, then } 
\begin{equation*}
\lim_{\lambda \rightarrow 0}\langle \Phi _{Q}^{F},B_{Q}^{F\dagger
}(\sum_{\omega \in F}x_{\omega }^{(1)}\lambda \int_{S_{\omega
}^{(1)}}^{T_{\omega }^{(1)}}S_{u_{1}}^{\omega }f_{\omega
}^{(1)}du_{1})...B_{Q}^{F\dagger }(\sum_{\omega \in F}x_{\omega
}^{(n)}\lambda \int_{S_{\omega }^{(n)}}^{T_{\omega }^{(n)}}S_{u_{n}}^{\omega
}f_{\omega }^{(n)}du_{n})\ \Phi _{Q}^{F}\rangle
\end{equation*}
\textit{exists uniformly for the }$x$\textit{'s and }$[S,T]$\textit{'s in a
bounded set of }$R$\textit{\ and is equal to } 
\begin{equation*}
\langle \Phi _{R}^{Q},W(\oplus _{\omega \in F}(x_{\omega }^{(1)}\chi
_{\lbrack S_{\omega }^{(1)},T_{\omega }^{(1)}]}\otimes f_{\omega
}^{(1)}))...W(\oplus _{\omega \in F}(x_{\omega }^{(n)}\chi _{\lbrack
S_{\omega }^{(n)},T_{\omega }^{(n)}]}\otimes f_{\omega }^{(n)}))\ \Phi
_{R}^{Q}\rangle .
\end{equation*}

For a proof see [10].

\bigskip 

{\large APPENDIX\ C:}\textbf{\ The Quantum Stochastic Differential Equation
for $U_{t}$.}

For convenience we shall consider only one coupling frequency $\omega $ so
that 
\begin{equation*}
H_{I}\equiv i\hbar \sum_{j}(D_{j}\otimes A^{\dagger }(g_{j}))
\end{equation*}
and we have dropped the superscript $\omega $ from the operators. Also we
shall consider only the Fock (vacuum) case $Q=1$. We define $G_{\lambda }(t)$
by 
\begin{equation*}
\langle \psi ,G_{\lambda }(t)\rangle =\mathrm{Tr}\{|\phi \rangle \langle
\psi |\otimes B_{T^{\prime }/\lambda ^{2}}^{(\omega ,\lambda )\dagger
}(f^{\prime })|\Psi _{R}\rangle \langle \Psi _{R}|B_{T/\lambda
^{2}}^{(\omega ,\lambda )}(f)U_{t/\lambda ^{2}}^{(\lambda )}\},\ 
\end{equation*}
where $\psi ,\phi \in \mathcal{H}_{S}$ and the second inner product is meant
on $\mathcal{H}_{S}\otimes \Gamma _{B}(L^{2}(\mathbb{R},K_{\omega }))$: we
know that the limit $\lim_{\lambda \rightarrow 0}\langle \psi ,G_{\lambda
}(t)\rangle $ exists and equals 
\begin{equation*}
\langle \psi \otimes B_{Q}^{\omega \dagger }(f,T)\Phi _{Q}^{\omega
},U_{t}\phi \otimes B_{Q}^{\omega \dagger }(f^{\prime },T^{\prime })\Phi
_{Q}^{\omega }\rangle .\ 
\end{equation*}
It is easy to show that this limit has the form $\langle \psi ,G(t)\rangle $
where $:t\mapsto G(t)\in \mathcal{H}_{S}$ is weakly differentiable. In order
to obtain a differential equation for $G(t)$ we note that for fixed $\lambda 
$ one has 
\begin{equation*}
\end{equation*}
\begin{eqnarray*}
{\dfrac{d}{dt}}\langle \psi ,G_{\lambda }(t)\rangle &=&\mathrm{Tr}\{|\phi 
\rangle \langle \psi |\otimes B_{T^{\prime }/\lambda ^{2}}^{(\omega
,\lambda )\dagger }(f^{\prime })|\Psi _{R}\rangle \langle \Psi
_{R}|B_{T/\lambda ^{2}}^{(\omega ,\lambda )}(f)\, 
{\dfrac{1}{\lambda }}\sum_{j}(D_{j}\otimes A^{\dagger }(S_{t/\lambda
^{2}}^{\omega }g_{j})-D_{j}^{\dagger }\otimes A(S_{t/\lambda ^{2}}^{\omega
}g_{j}))U_{t/\lambda ^{2}}^{(\lambda )}\} \\
&=&\Gamma _{\lambda }+\Xi _{\lambda };
\end{eqnarray*}
\begin{equation*}
\ 
\end{equation*}
where 
\begin{eqnarray*}
\Gamma _{\lambda } &=&{\dfrac{1}{\lambda }}\mathrm{Tr}\{|\phi \rangle
\langle \psi |\otimes B_{T^{\prime }/\lambda ^{2}}^{(\omega ,\lambda
)\dagger }(f^{\prime })|\Psi _{R}\rangle \langle \Psi _{R}|B_{T/\lambda
^{2}}^{(\omega ,\lambda )}(f)\,\sum_{j}D_{j}\otimes A^{\dagger
}(S_{t/\lambda ^{2}}^{\omega }g_{j})U_{t/\lambda ^{2}}^{(\lambda )}\}, \\
\Xi _{\lambda } &=&-{\dfrac{1}{\lambda }}\mathrm{Tr}\{|\phi \rangle \langle
\psi |\otimes B_{T^{\prime }/\lambda ^{2}}^{(\omega ,\lambda )\dagger
}(f^{\prime })|\Psi _{R}\rangle \langle \Psi _{R}|B_{T/\lambda
^{2}}^{(\omega ,\lambda )}(f)\,\sum_{j}D_{j}^{\dagger }\otimes
A(S_{t/\lambda ^{2}}^{\omega }g_{j})U_{t/\lambda ^{2}}^{(\lambda )}.
\end{eqnarray*}
\begin{equation*}
\end{equation*}
Now 
\begin{eqnarray*}
\Gamma _{\lambda } &=&\sum_{j}{\dfrac{1}{\lambda }}\lambda
\int_{0}^{T/\lambda ^{2}}\langle S_{u}^{\omega }f,S_{t/\lambda ^{2}}^{\omega
}g_{j}\rangle du\langle D_{j}^{\dagger }\psi ,G_{\lambda }(t)\rangle  \\
&=&\sum_{j}{\dfrac{1}{\lambda }}\lambda \int_{(-t)/\lambda
^{2}}^{(T-t)/\lambda ^{2}}\langle S_{v}^{\omega }f,g_{j}\rangle dv\langle
D_{j}^{\dagger }\psi ,G_{\lambda }(t)\rangle 
\end{eqnarray*}
where we made the substitution $u-t/\lambda ^{2}=v$. We see that for bounded 
$D$ this converges as $\lambda \rightarrow 0$ a.e. to 
\begin{equation*}
\sum_{j}\chi _{\lbrack S,T]}(f|g_{j})_{\omega }\langle D_{j}^{\dagger }\psi
,G_{\lambda }(t)\rangle .
\end{equation*}
Next of all the term $\Xi _{\lambda }$ must be reordered as follows 
\begin{eqnarray*}
\Xi _{\lambda } &=&-{\dfrac{1}{\lambda }}\sum_{j}\mathrm{Tr}\{|\phi \rangle
\langle \psi |\otimes B_{T^{\prime }/\lambda ^{2}}^{(\omega ,\lambda
)\dagger }(f^{\prime })|\Psi _{R}\rangle \langle \Psi _{R}|B_{T/\lambda
^{2}}^{(\omega ,\lambda )}(f)\,(D_{j}^{\dagger }\otimes 1)U_{t/\lambda
}^{(\lambda )}(1\otimes A(S_{t/\lambda ^{2}}^{\omega }))\} \\
&&-{\dfrac{1}{\lambda }}\sum_{j}\mathrm{Tr}\{|\phi \rangle \langle \psi
|\otimes B_{T^{\prime }/\lambda ^{2}}^{(\omega ,\lambda )\dagger }(f^{\prime
})|\Psi _{R}\rangle \langle \Psi _{R}|B_{T/\lambda ^{2}}^{(\omega ,\lambda
)}(f)\,(D_{j}^{\dagger }\otimes 1)[(1\otimes A(S_{t/\lambda ^{2}}^{\omega
})),U_{t/\lambda }^{(\lambda )}]\} \\
&=&\Xi _{\lambda }^{a}+\Xi _{\lambda }^{b}.
\end{eqnarray*}
In a fashion similar to the calculation of $\Gamma _{\lambda }$, one easily
arrives at 
\begin{equation*}
\lim_{\lambda \rightarrow 0}\Xi _{\lambda }^{a}=-\sum_{j}\chi _{\lbrack
0,T^{\prime }]}(g_{j}|f^{\prime })_{\omega }\langle D_{j}^{\dagger }\psi
,G_{\lambda }(t)\rangle ,\mathrm{a.e.}
\end{equation*}
To evaluate the limit of $\Xi _{\lambda }^{b}$ we note that 
\begin{eqnarray*}
[(1\otimes A(S_{t/\lambda ^{2}}^{\omega }g_{j})),U_{t/\lambda
^{2}}^{(\lambda )}] 
&=&\sum_{n=1}^{\infty }({\dfrac{\lambda }{i\hbar }})^{n}\int_{0}^{t/\lambda
^{2}}dt_{1}\int_{0}^{t_{1}}dt_{2}...\int_{0}^{t_{n-1}}dt_{n}[(1\otimes
A(S_{t/\lambda ^{2}}^{\omega }g_{j})),H_{I}(t_{1})...H_{I}(t_{n})] \\
&=&\sum_{n=1}^{\infty }({\dfrac{\lambda }{i\hbar }})^{n}\int_{0}^{t/\lambda
^{2}}dt_{1}\int_{0}^{t_{1}}dt_{2}...\int_{0}^{t_{n-1}}dt_{n}\{[(1\otimes
A(S_{t/\lambda ^{2}}^{\omega
}g_{j})),H_{I}(t_{1})]H_{I}(t_{2})...H_{I}(t_{n}) \\
&&+H_{I}(t_{1})[(1\otimes A(S_{t/\lambda
^{2}}g_{j}),H_{I}(t_{2})...H_{I}(t_{n})]\}.
\end{eqnarray*}
It can be shown that in the limit $\lambda \rightarrow 0$ only the
commutator involving $H_{I}(t_{1})$ contributes. Hence 
\begin{equation*}
\lim_{\lambda \rightarrow 0}\Xi _{\lambda }^{b}=-\lim_{\lambda \rightarrow
0}\sum_{n=1}^{\infty }({\dfrac{\lambda }{i\hbar }})^{n-1}\int_{0}^{t/\lambda
^{2}}dt_{1}\int_{0}^{t_{1}}dt_{2}...\int_{0}^{t_{n-1}}dt_{n}
\end{equation*}
\begin{equation*}
\sum_{j,k}\langle S_{t/\lambda ^{2}}^{\omega }g_{j},S_{t_{1}}^{\omega
}g_{k}\rangle \mathrm{Tr}\{|\phi \rangle \langle \psi |\otimes B_{T^{\prime
}/\lambda ^{2}}^{(\omega ,\lambda )\dagger }(f^{\prime })|\Psi _{R}\rangle
\langle \Psi _{R}|B_{T/\lambda ^{2}}^{(\omega ,\lambda )}(f)\,D_{j}^{\dagger
}D_{k}H_{I}(t_{2})...H_{I}(t_{n})\},
\end{equation*}
but this is the same as 
\begin{eqnarray*}
&&-\lim_{\lambda \rightarrow 0}{\dfrac{1}{\lambda ^{2}}}\sum_{j,k}%
\int_{0}^{t}ds\langle S_{t/\lambda ^{2}}^{\omega }g_{j},S_{s/\lambda
^{2}}^{\omega }g_{k}\rangle D_{j}^{\dagger }D_{k}U_{s/\lambda
^{2}}^{(\lambda )} \\
&=&-\lim_{\lambda \rightarrow 0}{\dfrac{1}{\lambda ^{2}}}\sum_{j,k}%
\int_{0}^{t}ds\langle S_{t/\lambda ^{2}}^{\omega }g_{j},S_{s/\lambda
^{2}}^{\omega }g_{k}\rangle \langle D_{k}^{\dagger }D_{j}\psi ,G_{\lambda
}(t)\rangle =-\sum_{j,k}(g_{j}|g_{k})^{\omega -}\langle \psi ,D_{j}^{\dagger
}D_{k}G(t)\rangle ,
\end{eqnarray*}
where we have used a technical lemma (6.3) of [1].

We now have 
\begin{eqnarray*}
\langle \psi ,G(t)\rangle  &=&\lim_{\lambda \rightarrow 0}\langle \psi
,G_{\lambda }(t)\rangle =\langle \psi ,G(0)\rangle +\lim_{\lambda
\rightarrow 0}\int_{0}^{t}(\Gamma _{\lambda }(s)+\Xi _{\lambda }(s))ds \\
&=&\langle \psi ,G(0)\rangle +\int_{0}^{t}ds\{\sum_{j}\chi _{\lbrack
S,T]}(s)(f|g_{j})_{\omega }\langle D_{j}^{\dagger }\psi ,G(s)\rangle  \\
&&-\sum_{j}\chi _{\lbrack S^{\prime },T^{\prime }]}(s)(g_{j}|f)^{\omega
}\langle \psi ,D_{j}G(s)\rangle -\sum_{j,k}(g_{j}|g_{k})^{\omega -}\langle
\psi ,D_{j}^{\dagger }D_{k}G(s)\rangle \}.
\end{eqnarray*}
Here we have written $(g|f)^{\omega -}$ for $(g|f)_{Q+}^{\omega -}$ when $Q=1
$.

The quantum stochastic differential equation corresponding to this integral
equation is 
\begin{equation*}
dU_{t}=\{\sum_{j}(D_{j}\otimes dB_{Q}^{\omega \dagger
}(g_{j},t)-D_{j}^{\dagger }\otimes dB_{Q}^{\omega
}(g_{j},t))-\sum_{j,k}(g_{j}|g_{k})^{\omega -}D_{j}^{\dagger }D_{k}dt\}\
U_{t}.
\end{equation*}
The generalization to $Q > 1$ and several coupling frequencies $\omega $
is now obvious.


\begin{thebibliography}{99}
\bibitem{1}  Accardi,L. ,Frigerio,A. and Lu, Y.G.:The weak coupling limit as
a quantum functional central limit. Comm.Math.Phys. 131,537-570(1990) .2cm

\bibitem{2}  Louisell,W. :Quantum statistical properties of radiation. John
Wiley and Sons.(1973).

\bibitem{3}  Pul\'{e},J.V. :The Bloch equations. Comm.Math.Phys.
38,241-256(1974)

\bibitem{4}  Davies,E.B. :Markovian master equation. Comm.Math.Phys.
39,91-110(1974)

\bibitem{5}  Haken,H. :Laser Theory. Springer(1984)

\bibitem{6}  Lax,M. :Phys.Rev. 145,111-129(1965)

\bibitem{7}  von Waldenfels,W. :Ito solution of the linear quantum
stochastic differential equation describing light emission and absorption.
SLNM 1055 (ed Accardi, Frigerio, Gorini)

\bibitem{8}  Accardi,L. ,Frigerio,A. and Lu, Y.G.:The weak coupling limit
(II): The Langevin equation and finite temperature case. Preprint Volterra
Centro No.13 (1989)

\bibitem{9}  Accardi,L. ,Frigerio,A. and Lu, Y.G.:The weak coupling limit
for Fermions. Preprint Volterra Centro No.12 (1989)

\bibitem{10}  Accardi,L. ,Frigerio,A. and Lu, Y.G.:The weak coupling limit
without rotating wave approximation. Preprint Volterra Centro No.23 (1990)

\bibitem{11}  Accardi,L. ,Frigerio,A. and Lu, Y.G.:Unified Approach to the
Quantum Master and Langevin Equations. Preprint Volterra Centro No.69 (1991)

\bibitem{12}  Accardi,L. and Lu, Y.G.:On the weak coupling limit for quantum
electrodynamics. Prob. Meth. in Math. Phys. (ed F.Guerra, M.I. Loffredo, C.
Marchro) World Scientific (1992) pgs 16-22.

\bibitem{13}  Hudson,R.L. , Parthasarathy,K.R. :Quantum Ito's formula and
stochastic evolutions. Comm.Math. Phys. 131,537-570 (1990)

\bibitem{14}  Parthasarathy,K.R.: An introduction to Quantum Statistical
Calculus. Monographs in Mathematics. Birkhauser (1992)

\bibitem{15}  Kittel,G. ,Quantum theory of solids, John Wiley and Sons(1963)

\bibitem{16}  N.N.Bogolubov and N.N.Bogolubov Jnr. Some applications of the
Polaron Theory. World Scientific Lecture Notes in Mathematics, vol 4 (1992)

\bibitem{17}  Sewell, G.L., Quantum Theory of Collective Phenomena,
Monographs on the physics and chemiistry of materials, Oxford Science
Publications. (1986) .2cm

\bibitem{18}  Collett, M.J., Gardiner, C.W.: Phys.Rev A 1386 (1984)

\bibitem{19}  Messiah, A.: Quantum Mechanics. Vol II, North Holland
Publishing Company. (1961)

\bibitem{20}  Au, C-K. and Feinberg,G.: Phys. Rev. A 9 ,1974 (1974) and
12,1772 (1975)

\bibitem{21}  Ford,G.W. von Waldenfels,W.: Radiative Energy Shifts for a
Nonrelativistic Atom.
\end{thebibliography}
\end{document}